\documentclass[12pt]{article}
\usepackage{color,soul}  
\usepackage{tikz} 
\usepackage{url}
\usepackage{footmisc}
\usepackage{longtable}
\usepackage{amsmath,amsfonts,amssymb,amsthm}
\usepackage{mathtools}
\usepackage{enumerate}
\usepackage{siunitx}
\usepackage{amsmath,amssymb,amsthm}
\usepackage{graphicx}
\usepackage{placeins}
\usepackage{bigstrut}
\usepackage{indentfirst}
\usepackage{natbib}
\usepackage{verbatim}
\usepackage{multirow}
\usepackage{epstopdf}
\usepackage{adjustbox}
\usepackage[titletoc,title]{appendix}
\usepackage{setspace}
\usepackage{bbm,bm}
\usepackage[left=1.00in, right=1.00in, top=1.00in, bottom=1.00in]{geometry}
\usepackage{color, colortbl}
\usepackage{import}
\usepackage{bbding}
\usepackage{multicol}
\usepackage[labelfont=bf,textfont=bf,]{caption}
\usepackage{graphicx, graphics}
\usepackage{booktabs, threeparttable}
\usepackage{rotating}
\usepackage{subcaption}
\usepackage{soul}
\usepackage[T1]{fontenc}

\usepackage{pifont}
\newcommand{\xmark}{\ding{55}}%

\newcommand{\sym}[1]{#1} 

\newtheorem{hyp}{Hypothesis}

\definecolor{sangre}{rgb}{0.6,0.18,0.19}
\definecolor{dullmagenta}{rgb}{0.4,0,0.4}
\definecolor{darkblue}{rgb}{0,0,0.6}
\ifx\pdfoutput\undefined
\usepackage[backref=none,hypertex,colorlinks,urlcolor = darkblue,citecolor = sangre,linkcolor=blue]{hyperref}
\else
\usepackage[backref=none,pdftex,hypertexnames=false,colorlinks,urlcolor = darkblue,citecolor =  sangre,linkcolor=blue]{hyperref}
\usepackage{scalerel,stackengine}
\stackMath
\newcommand\reallywidehat[1]{%
  \savestack{\tmpbox}{\stretchto{%
      \scaleto{%
        \scalerel*[\widthof{\ensuremath{#1}}]{\kern-.6pt\bigwedge\kern-.6pt}%
        {\rule[-\textheight/2]{1ex}{\textheight}}
      }{\textheight}%
    }{0.5ex}}%
  \stackon[1pt]{#1}{\tmpbox}%
}
\usepackage{enumitem} 
\usepackage{pifont}

\makeatletter
\newcommand*\bigcdot{\mathpalette\bigcdot@{.5}}
\newcommand*\bigcdot@[2]{\mathbin{\vcenter{\hbox{\scalebox{#2}{$\m@th#1\bullet$}}}}}
\newcommand\EightPtClose{\@setfontsize\EightPtClose\@viiipt{9}}
\newcommand\TenPtType{\@setfontsize\TenPtType\@xpt\@xiipt}
\def\notesize{\TenPtType}
\def\notesize{\EightPtClose}
\newenvironment{figurenotes}[1][Note]{\begin{minipage}[t]{\linewidth}\notesize{\itshape#1: }}{\end{minipage}}
\makeatother

\usepackage{array}
\newcolumntype{P}[1]{>{\centering\arraybackslash}p{#1}}

\usepackage{amsmath}
\makeatletter
\newcommand*\rel@kern[1]{\kern#1\dimexpr\macc@kerna}
\newcommand*\widebar[1]{%
  \begingroup
  \def\mathaccent##1##2{%
    \rel@kern{0.8}%
    \overline{\rel@kern{-0.8}\macc@nucleus\rel@kern{0.2}}%
    \rel@kern{-0.2}%
  }%
  \macc@depth\@ne
  \let\math@bgroup\@empty \let\math@egroup\macc@set@skewchar
  \mathsurround\z@ \frozen@everymath{\mathgroup\macc@group\relax}%
  \macc@set@skewchar\relax
  \let\mathaccentV\macc@nested@a
  \macc@nested@a\relax111{#1}%
  \endgroup
}

\makeatother
\usepackage[capitalize]{cleveref}

\begin{document}

\title{\Large 
Security Issuance, Institutional Investors, and Quid Pro Quo\thanks{\protect\linespread{1}\protect\selectfont  We would like to thank Christian Opp, Joanna Wu, Michael Gofman, and Bill Wilhelm for their helpful comments and suggestions. We thank participants and discussants at several conferences for their comments. We also thank the 3Cavaliers at the University of Virginia for their generous research support and Lam Bui and Hengxiang Cao for their excellent research assistance.}
}

\author{Gaurab Aryal\thanks{Department of Economics, Boston University, \href{mailto:aryalg@bu.edu}{aryalg@bu.edu}.} \qquad
Zhaohui Chen\thanks{McIntire School of Commerce, University of Virginia, \href{mailto:zc8j@virginia.edu}{zc8j@virginia.edu}.}\qquad
Yuchi Yao\thanks{Lundquist College of Business, University of Oregon, \href{mailto:yuchiyao@uoregon.edu}{yuchiyao@uoregon.edu}}\qquad
Chris Yung\thanks{McIntire School of Commerce, University of Virginia, \href{mailto:cay2m@virginia.edu}{cay2m@virginia.edu}}
}

\date{\today}

\maketitle
\thispagestyle{empty}

\begin{abstract}

Securities issuance through intermediaries is subject to agency problems and informational frictions. We examine these effects using SPAC data. We identify ``premium'' investors whose participation is linked to lower liquidation risk, higher returns, and lower redemption rates, consistent with both informational rents and agency frictions. In contrast, ``non-premium'' investors engage in non-agency \emph{quid pro quo} relationships. Specifically, they receive high returns from an intermediary (quid) in exchange for a tacit agreement to participate in weaker future deals (quo). These relationships serve as insurance for issuers and intermediaries, enabling more issuers to access markets.

\bigskip

\noindent\textbf{Keywords: Security Issuance, Institutional Investors, SPAC, Underpricing}

\end{abstract}

\newpage
\clearpage
\setcounter{page}{1}
\section{Introduction}

One common view is that IPO issuers face an agency problem in which underwriters do not maximize sale proceeds but instead engage in quid pro quo arrangements with investors. Specifically, underwriters offer low-priced shares to favored investors, expecting to be repaid in other transactions \citep{RitterWelch2002} benefiting the underwriter rather than the issuer. Drawing upon earlier work,\footnote{See, for example, \cite{BenvenisteBusaba} and \cite{BenvenisteLjungqvistWilhelmYu2003}.} \cite{LJUNGQVIST2007375} posits a second type of quid pro quo:
\begin{quote}
 A second advantage of repeated interaction is that it allows underwriters to ‘bundle’ offerings across time. To ensure continued access to lucrative IPOs in the future, investors will occasionally buy poorly received IPOs as long as the loss they suffer in any given IPO does not exceed the present value of future rents they expect to derive from doing business with the underwriter.   
\end{quote}

Under the first view (``agency theory''), quid pro quo causes a deadweight loss in the securities market. It acts as a tax on issuers, decreasing their incentive to raise capital. By contrast, under the second view, quid pro quo arrangements represent a transfer from high-quality to low-quality issuers. For issuers uncertain about their valuation, this arrangement can ex-ante be helpful. It acts as insurance, allowing them to raise capital even when given a lukewarm market reception. Under this view (``insurance hypothesis''), quid pro quo plays a central and beneficial role in the intermediation process.

In this paper, we provide evidence consistent with the insurance hypothesis. In doing so, our paper is the first paper to show evidence of a positive role of quid pro quo arrangements in intermediated finance. Although this theory should apply to any setting of security issuance, in this paper, we focus on special purpose acquisition companies (SPACs), which are publicly traded shell companies formed by sponsors to merge with private targets, thereby taking them public. Unlike standard IPO data, SPAC data have unique features (that we make precise shortly below) that allow us to observe quid pro quo arrangements directly. 

In our analysis, the sponsor (rather than a bank) is the intermediary. Like underwriters in a traditional IPO, sponsors serve as intermediaries between buyers (investors) and sellers (the issuer). In both cases, an intermediary negotiates with the issuer to determine the offer price and number of shares sold and helps determine the share allocation across investors. In both cases, the intermediary profits only if the deal is completed.

The identification challenge we face is that investors may perform dual roles--they provide information while also engaging in quid pro quo arrangements. For example, suppose a SPAC stock price rises upon the announcement of a particular investor's involvement. This increase could be due to investor certification or the quid pro quo relationship, where the investor lacks private information but is allocated hot deals. The latter also leads to a price increase contemporaneously with the allocation, which is unrelated to certification.

We discuss these challenges in more detail later, but to briefly preview our results, we conclude that it is difficult to separate these two phenomena. This observation echoes the sentiment of \cite{lowry2017initial} in their IPO survey paper, who lament the lack of progress on this separation, stating: 

\begin{quote}
“Does [IPO underpricing] reflect a reward to institutions for sharing value-relevant information with underwriters, or does it reflect a quid pro quo arrangement where the bank is rewarding its best clients? [..] This is an immensely important issue and one that we hope future research will be able to address.”
\end{quote}

By contrast, as we make clear in our hypothesis design section, the insurance hypothesis makes distinct predictions from the other theories, and our contribution is to isolate its effects in our data.

To this end, we distinguish between two types of investors: ``premium'' and ``non-premium'' (defined shortly below). We find evidence consistent with premium investors acting \emph{only} in either the certification or agency role. Their presence is associated with higher announcement period returns, less redemption, and a higher likelihood of a successful SPAC merger. By contrast, non-premium investors engage \emph{only} in quid pro quo arrangements as described by the insurance hypothesis. Specifically, we track the returns in each sponsor-investor pair, showing that within each pair, returns are mean-reverting. This pattern is consistent with the buildup and drawdown of relationship capital as per our insurance hypothesis. Second, we show that when a deal is weak, sponsors allocate higher shares to non-premium investors who earned higher returns in the past from the same sponsors than other non-premium investors who have yet to work with those sponsors. Third, the buildup of goodwill in this fashion serves as insurance by reducing the likelihood that a current SPAC is liquidated.  

To summarize, the insurance hypothesis form of quid pro quo is evident in our data, and this role is played only by non-premium investors. Further, it is not destructive to issuers but enables a wider range of issuers to access public markets.  

As mentioned before, the SPAC setting offers several advantages in terms of research design over the traditional IPO. First, a SPAC divides the process of going public into distinct phases. In the initial stage, the sponsor raises capital from investors. A SPAC starts as a newly formed \emph{shell company} without an operating business, so there is limited scope for information asymmetry (about the target firm) compared to a traditional IPO that starts with an operating business. The next phase consists of a ``business combination'' in which the SPAC merges with a private business and introduces new institutional investors known as ``PIPE” (private investment in public equity), allowing us to isolate their effects on SPACs.

Second, legally, SPACs are publicly traded firms. Consequently, we observe the allocation to each PIPE investor and the effect of major events (e.g., business combination) on share prices. These data enable us to measure the return history for every sponsor-PIPE pair and create an in-sample measure of past relationships.\footnote{In traditional IPOs, allocations to institutional investors are unreported, and the quarterly 13F filings are insufficient to quantify favoritism if preferred clients ``flip'' shares in the secondary market. Furthermore, 13F filings do not distinguish shares allocated at the IPO from shares purchased in the secondary market.} 

Our data include the universe of (1,072) SPACs publicly listed from January 1, 2010, to December 31, 2021.
We follow these SPACs and record all major events and market returns until February 17, 2023. 
To categorize investor types, we follow the argument of \cite{benveniste1989investment}, who argue that all else equal, frequent and large investors are more likely to produce value-relevant information than others.  To directly implement this notion, we classify PIPE investors into premium and non-premium types based on participation frequency and investment size using \emph{k-means clustering}. We refer to such investors as premium investors and estimate their effects on SPAC outcomes.\footnote{We consider an alternative approach to assess if the insurance hypothesis is sensitive to how we define premium investors. For each PIPE investor, we searched and counted the articles published in The New York Times between January 2010 and December 2021 that featured the investor. We then define the investors with the top 5\% counts as premium investors. This alternative definition also supports the insurance hypothesis.}
  
A challenge in estimating the effects of premium investors is that the set of investors working with a sponsor is non-random. Better or more experienced sponsors may choose to work with better premium investors. If so, the estimated effects of premium investors could be biased because of the \emph{assortative matching}. To address this issue, we first estimate fixed effects for each sponsor and investor as proxies for their (unobserved) qualities following the methodology of \cite{abowd1999high} and then use those fixed effects in our regression analysis as controls. Our identification assumption is that conditional on those fixed effects, the sponsor and premium investor matching is random.

We have several interesting results that shed light on the workings of the securities market. First, we find evidence consistent with premium investors offering value-relevant information and playing certification roles. 
Their participation is associated with significantly less share redemption by public investors and a large positive announcement effect on the stock price. In particular, one standard deviation (32.85 pps) increase in premium investors’ participation is associated with a 5.91 pps decrease in the redemption rate (10.5\% of the sample mean) and a 4.40 pps increase in the announcement-day return (117.7\% of the sample mean). 
  
Sponsors profit an average of \$126.9 million when the merger is successful; otherwise, they lose their initial investment, which averages \$8.2 million. Therefore, the sponsors benefit from investors' participation, which improves the chances of a successful merger. In this sense, the quid pro quo arrangement suggested by \cite{LJUNGQVIST2007375} may benefit not only issuers in the form of insurance but also the sponsors. 
Thus, sponsors may have incentives to offer discounted shares to those investors in strong deals in exchange for an implicit promise to help them out with weaker deals in the future. We find evidence that \emph{only} the non-premium investors engage in such quid pro quo arrangements.  

For our second set of findings, we use the instrumental variables approach to document three ways quid pro quo manifests in the data. First, for each sponsor and non-premium investor pair, money left on the table (``MLOT'') exhibits mean-reversion, i.e., non-premium investors profit less in the current deal when they have high returns on past deals, and vice versa. In particular, 
one dollar \emph{above} average MLOT in the last deal leads to a \$2.72 \emph{below} average MLOT from the current deal. 
We interpret this \emph{mean-reversion} as the sponsor building a relationship with some non-premium investors and drawing it down when needed. 

Second, when the deal is weak or ``cold,'' i.e., when the offer share price at IPO is higher than the price on the closing day of the merger, the sponsor allocates shares disproportionately to those who earned strong returns from \emph{that same} sponsor in the past. 
Moreover, the sponsor allocates more securities to non-premium investors with whom it has a strong relationship when the deals are weak than when the deals are strong.

Third, a sponsor with strong relationships with non-premium investors is more likely to succeed. More precisely, if sponsors allocate one standard deviation more MLOT above average (\$300 million) to non-premium investors, the liquidation probability of the current deal decreases by 48.93 pps. As a reference, the average liquidation rate in our sample is 17.82\%, so this decrease is economically meaningful. Using several robustness exercises, including placebo exercises, we verify that our results are consistent with how we define premium and non-premium investors.

Thus, we establish a new institutional feature in security issuance based on the relationship between the intermediary and institutional investors. A premium investor is paid on average \$6.96 million per deal, which, our estimates suggest, is a payment to produce information for the current target. A non-premium investor is paid, on average \$3.17 million to reduce liquidation risk if, in the future, the current sponsor wants to merge with a weak target. Thereby, we provide evidence that the quid pro quo arrangement enables cross-deal subsidies from high-quality targets to low-quality targets that help the latter to go public. 
In contrast, the existing literature treats quid pro quo as a pure agency cost for the issuers.

We contribute to a long line of research that uses underpricing to infer the importance of agency and informational problems. Underpricing has been argued to be an implication of information asymmetry between investors \citep{ROCK1986} or between investors and intermediaries \citep{benveniste1989investment}. In contrast, \cite{BeattyRitter1986, LoughranRitter2002} and \cite{LoughranRitter2004} argue that underpricing is pure unproductive rent. One debate has been over the nature of `partial adjustment,' i.e., underwriters partially adjust to good information revealed by investors but fully adjust to bad information. \cite{hanley1993underpricing} presents supporting empirical evidence. Also see \cite{BradleyJordan2002} and \cite{Ince} for more. 
We complement this literature by examining broader outcomes and showing that information and quid pro quo arrangements are present. 

Our approach is partly motivated by the view of \cite{BenvenisteBusaba} that underwriters strategically ``bundle'' IPOs to resolve information asymmetries among companies with similar technologies. They suggest that ``leaders'' in this context can pave the way for "followers" through information production, leading to a standoff where no company wants to go public first. This issue is mitigated by transfers between leaders and followers, facilitating waves of IPOs. \cite{BenvenisteLjungqvistWilhelmYu2003} support this theory by showing that while initial returns and IPO volume are positively correlated overall, this correlation turns negative among offerings influenced by a common valuation factor. Unlike these papers, our observation is at the sponsor-investor pair level. Using information on investor allocations, we provide evidence of relationships between individual actors in the IPO market.

Our findings about the differential effects of premium and non-premium investors complement the findings in \cite{WangYung2011} and  \cite{ChemmanueHuHuang2010}. 
Additionally, our finding that information production coexists with quid pro quo is reminiscent of \cite{JenkJones}. Using data on the initial allocations of investment banks participating in traditional IPOs, they find evidence that investors are rewarded more when they submit price-limited bids and participate more actively in the meetings before or during bookbuilding. Both behaviors are consistent with an informational role. However, they also find that allocations are affected by how much revenue the investor generates for the bank, which is consistent with the quid pro quo argument. They do not study the effect of quid pro quo or document any bundling as we consider here.

We contribute to the emerging literature on SPACs, examining their role and mechanics. \cite{ritter2021spacs} explore the costs and benefits of SPACs for going public; \cite{gofman2022spacs} investigate conflicts of interest and harm from directors on multiple SPAC boards; and \cite{KlausnerOhlroggeRuan2022} study SPAC structures and costs. \cite{AltiCohn} addresses the adverse selection in choosing between traditional IPOs and SPACs, modeling how SPAC sponsors ``cream skim'' the best issues via costly search.

\section{A Primer on SPACs}
A SPAC is formed when a sponsor begins the SEC-mandated initial public offering process. It is, first and foremost, an IPO and must comply with all SEC IPO regulations. In addition, SPACs are subject to additional requirements as ``blank check companies,'' including how they use their escrow fund (described below). 
The SPAC aims to identify and acquire a private company to take it public. 

During their IPOs, SPACs offer securities to public investors called \emph{units}, typically priced at \$10 per unit.\footnote{One exception  is \emph{Periphas Capital Partnering Corporation}, which priced its unit at \$25.} 
A unit comprises one Class A common stock share and a proportion (e.g., 1, 1/2, 1/3) of one redeemable warrant. Initially, these units trade as a whole. However, usually around the 52$^{nd}$ calendar day following the IPO, owners have the option to either continue to hold units as a whole or trade shares and warrants separately.\footnote{After the separation, SPACs do not issue fractional warrants and only whole warrants trade. In other words, the number of warrants issued after separation equals the fraction rounded down to the nearest whole number. For example, if a unit contains one share and 1/3 of a warrant, then an investor with 1 unit receives no warrant, but an investor with 30 units receives ten warrants.} 
Once the merger deal between the SPAC and a target is closed, a warrant gives an investor the right to purchase additional shares of common stock. The strike price is typically set well above the original SPAC price; a common choice is \$11.50 per share. Any shares created in this manner automatically convert into ordinary shares after a merger. 

Sponsors purchase shares before SPACs go public.
These shares are called the ``founder shares'' or the sponsor's ``promote''. Unlike Class A common shares in the units owned by public investors, the sponsor's shares become worthless if the SPAC liquidates. 

A SPAC generally has 24 months from its IPO to identify a target company, negotiate a deal, and complete the merger. 
Failure to complete the merger on time leads to its liquidation and the SPAC must redeem all public shares and pay its investors with the funds held in the trust account, including any interest earned.


Once the SPAC locates a target, it uses proceeds from the trust fund for the merger. 
Until then, the trust fund can be used only to pay public shareholders who redeem their shares. 
The sponsor creates a ``risk capital'' to meet operating expenses by selling private placement warrants. These warrants are irredeemable when the SPAC liquidates and are subject to transfer restrictions. 


Next, SPAC shareholders vote on the deal, and the transaction occurs only if a majority votes in favor. Shareholders can also redeem their shares, typically having at least 20 business days before the merger is complete. The merger agreement with the target company requires that the SPAC have a minimum net worth or a certain amount of cash (per the negotiation) as a closing condition. 

Typically, the purchase price of the target firm exceeds the funds held by the SPAC. In such cases, the SPAC raises additional funds from various private investors. Because the SPAC is a publicly traded firm, these funds are called PIPEs. More rarely (12\% of our sample), the SPAC purchases the target without involving PIPEs.

Sometimes, the funds raised in the SPAC IPO are sufficient to purchase the target the sponsor has located. More typically, the sponsor needs to raise additional funds. 

PIPE investors are accredited institutional investors, e.g., hedge funds and mutual funds, that agree to purchase securities issued by a company. Such an agreement requires the company to file a resale registration statement (in the future) to allow investors to resell their securities. Compared to raising capital in the public market, PIPE investments tend to have lower transaction costs, and companies can disclose transaction details \emph{after} receiving commitments from PIPE investors. 

\begin{table}[t!]
  \centering
  \caption{Example of Key Events and Timeline for a SPAC}
    \begin{tabular}{lc}
    \toprule
    \textbf{Key Events} & \textbf{Dates} \\
    \midrule
    S-1 Filed and PIPE (pre) Announced    & 08-Jan-2021 \\
    IPO    & 29-Jan-2021 \\
    Deal Announced, PIPE (at) Announced    & 06-Jul-2021 \\
    Proxy Filed    & 12-Aug-2021 \\
    PIPE (post) Announced & 18-Jan-2021 \\
    Shareholder Vote & 24-Jan-2022 \\
    Deal Closed    & 25-Jan-2022 \\
    Liquidation Deadline & 02-Feb-2023 \\
    \bottomrule
    \end{tabular}%
  \label{tab:key_events}
  \begin{figurenotes} Key events and the dates for the SPAC \emph{CF Acquisition Corp V}.\end{figurenotes}
\end{table}%

We categorize PIPE transactions based on the timing of their commitment: before, at, and after the business combination. In particular, the sponsor can invite PIPE investors before the SPAC IPO to commit to making certain investments in connection with the business combination. The PIPE transactions formed at the time of the IPO but before a target is identified are called forward purchase agreements; henceforth, they are PIPE (pre).
The second type of PIPE transaction is announced along with the news about the business combination; henceforth, PIPE (at). The third type of PIPE transaction occurs after the target is announced but before the merger is completed; henceforth, PIPE (post). 
A SPAC can have any combination of these three types of PIPEs. For example, Table  \ref{tab:key_events} shows an example of SPAC, CF Capital Acquisition Corp V, which had all three PIPEs.

\begin{figure}[ht!]
 \centering
 \caption{Events timing for Traditional IPO and SPAC}
 \medskip
 \raggedright
\centering
\includegraphics[scale=0.58]{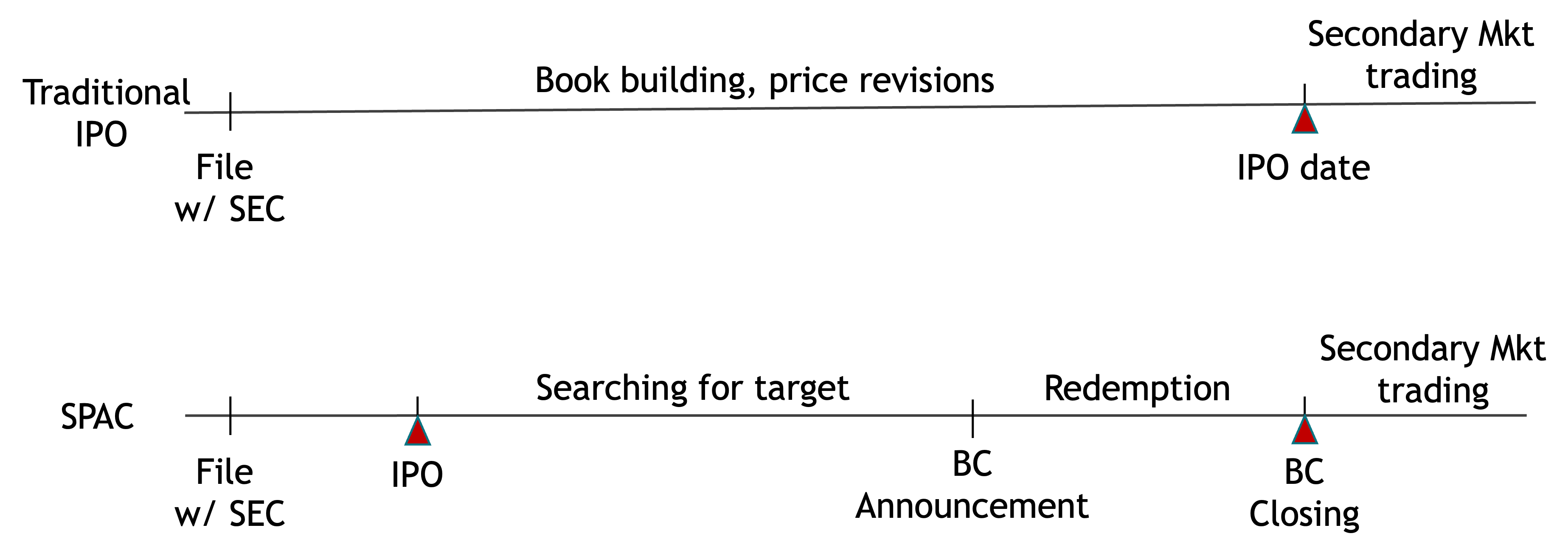}
\label{fig:timeline}
\begin{figurenotes}
This figure illustrates the timelines of a traditional IPO (top) and a SPAC (bottom). The acronym ``BC" stands for a business combination. 

\end{figurenotes}

\end{figure}

In Figure \ref{fig:timeline}, we compare the timeline of a SPAC and a traditional IPO. The date most analogous to the IPO date in a traditional initial public offering is the business combination closing: the point at which a private operating firm becomes public.

\section{Testable Hypotheses\label{section:hypotheses}}
\noindent We develop testable hypotheses based on the following theories:
\begin{itemize}
    \item Certification: reputable PIPE investors have private information and certify deals.
    \item Agency theory: Sponsors engage in quid pro quo arrangements with privileged PIPE investors at the expense of target firms.
    \item Insurance hypothesis:  Sponsors engage in quid pro quo arrangements with relationship PIPE investors in hot deals in order to complete future weaker deals.
\end{itemize}
\noindent We discuss the predictions of each paradigm in turn.

{\bf Certification Theory.} Under the first paradigm, investors have or produce private information about the issuer and earn rents accordingly. This paradigm views a central goal of the going public process as extracting value-relevant information about the issuer.

Going public via SPACs involves investors in three stages (see Figure \ref{fig:timeline}), and information production may have different implications at each step.   In particular, we predict zero underpricing during the SPAC IPO because a target has yet to be identified, and thus, there is no due diligence to perform. There is virtually no scope for investors to have private information; they exchange cash for a proportion of a trust fund of a fixed value.  In the business combination phase, investors participating in the target announcement (PIPE (at)) are the analogs to the buyers in IPO bookbuilding models. Specifically, the sponsor finds a target and approaches PIPE (at) investors to solicit interest. The sponsor builds a book of demand and--in conjunction with the selling firm--negotiates prices and allocations, considering any informational rents that need to be paid.

However, investors may differ in their abilities to produce information. 
For instance, \cite{benveniste1989investment} suggest that premium investors who frequently access the market are likely more informed than others. In our context, such investors have private information about the quality of the issuing (target) firm and overall conditions affecting the market valuation. Conversely, non-premium investors are less informed and consequently have less impact on prices. Under these assumptions, the securities of the target firm may be underpriced to compensate premium investors for truthfully disclosing their private information about the target. 

Furthermore, a premium PIPE (at) investor is incentivized to purchase more shares when its private information indicates the de-SPAC deal is profitable. Additionally, sponsors prioritize allocating to premium investors to encourage them to disclose their information. Consequently, a larger allocation to premium investors signals the good quality of the deal, which reduces redemption by public investors and leads to a positive market reaction. However, assuming non-premium PIPEs have no certification role, their participation affects neither redemption nor announcement-day returns.  

Conversely, we posit that PIPE (pre) and PIPE (post) investors do not play certification roles. In the case of PIPE (pre), there has yet to be any information about the target, so there is nothing to certify. PIPE (post) investors are used less often than PIPE (at) investors and can be used to fill the financing gap caused by redemption. If the sponsor intended a PIPE investor to be used for certification purposes (as opposed to ``gap filling''), the sponsor would have announced the participation of the PIPE contemporaneously when the deal was announced rather than waiting until after public investors reacted.  

Suppose a premium PIPE (at) investor acts purely as an information producer and is under no obligation to the sponsor to maintain a relationship. In this case, the PIPE participates in deals it likes and is absent otherwise. Such an arrangement should not generate any mean reversion when the unit of observation is sponsor-PIPE.  

\begin{hyp}Under the certification theory, the following should hold: 
\begin{enumerate}\label{hyp:info_ipo}
\item Allocations to premium PIPE (at) investors are associated with lower redemption and positive announcement-day returns.
\item For a sponsor-PIPE (at) relationship pair, MLOT is not mean-reverting.
\end{enumerate}
\end{hyp}

{\bf Agency Theory.} To develop predictions mirroring those of Hypothesis \ref{hyp:info_ipo}, we concentrate on the business combination phase, during which the sponsor acts as an intermediary between the selling firm (target) and investors. 
 Under this paradigm, investors lack private information. Instead, the sponsor acts as a self-interested intermediary, negotiating with the selling firm (target) to ensure that underpriced shares flow to privileged PIPE investors.  

Consistent with traditional IPO literature, we assume that the intermediary need not grant all investors the same level of privilege. Rather, the intermediary might only have relationships with certain investors allocated hot deals, whereas other (non-favored) investors receive fairly priced offerings. We refer to favored PIPEs as "relationship" investors and let the data determine their identity.

\begin{hyp}Under the agency theory, the following should hold: 
\begin{enumerate}\label{hyp:agency}
\item Allocations to relationship PIPE (at) investors are associated with lower redemption and positive announcement-day returns.
\item For a sponsor-PIPE (at) relationship pair, MLOT is not mean-reverting.
\end{enumerate}
\end{hyp}

Note that these predictions mirror Hypothesis \ref{hyp:info_ipo}, but the causality is reversed. Under Hypothesis \ref{hyp:info_ipo}, the SPAC stock price rises because the market believes that an informed investor has private information. Under Hypothesis \ref{hyp:agency}, the announcement of relationship investors coincides with the announcement of an underpriced target in a non-causal manner. Privileged investors receive underpriced shares but do not influence stock prices.

As pointed out in the introduction, the fact that these two theories make similar predictions makes it challenging to distinguish between them. The announcement of the target's identity coincides with the announcement of PIPE (at) investors, frustrating our ability to disentangle the effects of each announcement.

{\bf Insurance Hypothesis.} Under this paradigm, sponsors hope to offer a sequence of SPACs but cannot guarantee that every target they find is high-quality. To provide themselves insurance, when they obtain hot deals, they favor particular PIPE investors from whom they expect to return the favor.\footnote{Note that the enforcement of the implicit relationship agreement does not require a long history between a sponsor and an investor. Indeed, reneging on even one deal destroys relationship capital, assuming participation in a deal is publicly observable; see  \cite{FUDENBERG}.}  As with the agency theory, we are agnostic about which PIPE investors serve this relationship role but instead let the data decide.

Specifically, the sponsor benefits from creating goodwill among relationship investors, hoping to ``draw down'' on this goodwill in the future when the deals are weak. Consequently, we hypothesize that a SPAC's liquidation risk is negatively correlated with the amount of MLOT that relationship investors have received in the past from the sponsor of this SPAC. 

\begin{hyp}\label{hyp:insurance}
Under the insurance hypothesis, the following should hold:
\begin{enumerate}
\item Allocations to relationship investors are uncorrelated with redemption and announcement-day returns. \label{hyp:quid_redemption}
\item For a given sponsor-PIPE relationship pair, MLOT is mean-reverting.\label{hyp:quid_mlot}
\item When a deal is weak, the sponsor is more likely to allocate shares to relationship PIPE investors who earned large past MLOT.\label{hyp:quid_share}
\item SPACs are less likely to liquidate when the sponsor has earned goodwill through high past MLOT.\label{hyp:quid_liquidation}
\end{enumerate}
\end{hyp}

Hypotheses \ref{hyp:insurance}-2, \ref{hyp:insurance}-3, and \ref{hyp:insurance}-4 follow directly from the nature of relationships under this paradigm. To explain Hypothesis \ref{hyp:insurance}-1, note that sponsors in our sample may either build up or draw down relationship capital. On average, the presence of relationship investors is neither good nor bad news.

\section{Data}
We consider the universe of (1,072) SPACs publicly listed from January 1, 2010, to December 31, 2021.
We follow these SPACs and record all major events and market returns until February 17, 2023. 
See Figure \ref{fig:size_number_spacs} for the monthly scatter plot in Figure \ref{fig:size_number_spacs}, which includes the so-called boom and bust cycle for SPACs. We construct our sample by merging five datasets: (i) the \emph{SPAC Research},\footnote{\url{https://www.spacresearch.com/}.} (ii) the \emph{PrivateRaise}--SPACs Platform,\footnote{\url{https://www.privateraise.com/about/platform_spac.php}.} (iii) the \emph{PrivateRaise}--PIPEs Platform (iv) \emph{Bloomberg} data on SPAC prices, and (v) Super 8-K filings from SEC.\footnote{\url{https://www.sec.gov/edgar/search-and-access}.}$^{,}$\footnote{Thus, we exclude SPACs traded in the over-the-counter markets. After the merger between a SPAC and a target is complete, the SPAC is delisted. The new (merged) company may be assigned a new ticker symbol. We use Bloomberg-assigned unique identifiers to match across samples.}

The SPAC Research and PrivateRaise's SPACs platform databases contain information on SPACs, including important event dates. PrivateRaise's PIPE platform database includes comprehensive information on PIPE transactions. The platform covers more than 20,000 placement profiles of PIPE transactions. It also provides contract terms, investment amounts, types of transactions, prices, identities, and allocations of the PIPE investors.

For these 1,072 SPACs, we manually merged the first two databases. 
The issuer's name recorded in the PIPE transactions can be either that of SPAC or the post-merger company. We use both names to match the SPAC sample with the PrivateRaise database in those cases. We also use the total PIPE amount from PrivateRaise to cross-check the match to ensure match accuracy. If there are inconsistencies, we use the SPAC's SEC filings documents to record the correct amount. We then merge our sample with Bloomberg's pricing data; warrant prices are unavailable in CRSP. Our matching process yields detailed information on transactions for all three types of PIPE investments for each SPAC in our sample. 

In summary, we observe each SPAC's life-cycle status, IPO structure, IPO outcomes, sponsor and management team, redemption, merger details, merger outcome, PIPE transactions, and price movement across events. As of February 17, 2023, 473 of these 1,072 SPACs had successfully merged, 128 had announced that they had identified a target, 280 were in the process of identifying targets, and 191 had failed and liquidated.

The number of SPAC shares changes over time on account of i) investors exercising warrants, ii) redemption of shares, iii) securities purchased by PIPES, and iv) the fact that founders sometimes voluntarily forfeit securities. Consequently, the number of shares listed in the IPO prospectus is no longer current as of the business combination date. We use Super 8-K (Super 20-F for foreign issuer) filings to collect information on the number and types of outstanding securities on the first listing day of the post-merger company. Information about securities at the time of the SPAC IPO may not be comprehensive either because the sponsors have forfeited their founder shares and private placement securities or the units that contain rights would have become common shares or both. We manually collect information from the Super 8-K (or Super 20-F) filings to accurately measure the number of outstanding securities post-merger. SEC mandates that SPACs file Super 8-K within four business days after the merger is complete, and it records the shares and warrants held by sponsors and public investors.


Table \ref{tab:sum_stat} provides summary data.
On average, SPACs raise \$272.42 million proceeds during their IPO, with a median of \$240 million and 90th percentile at \$440 million. 
A typical SPAC unit contains 0.44 warrants and 0.01 rights. During the IPO process, SPACs' sponsors exercise the over-allotment option (on average) 11.59

SPACs take, on average, 361 days to find a target and an additional 168 days to complete the merger process. After targets' announcements, 56.54\% of SPAC's public shareholders redeem their shares on average.
Table \ref{tab:sum_stat} also displays information about the participation of PIPE investors for 473 SPACs that completed the merger process during our sample period. 
At the time of their IPOs, 10\% of SPACs have pre-committed investment from PIPE investors, i.e., PIPE (pre). 
When SPACs announce their targets, 66\% of them also announce investments from PIPE (at), and 12\% of SPACs obtain additional investments from PIPE (post) after the target announcement.

{\bf Returns.} 
As seen in Table \ref{tab:sum_stat}, the average return for SPAC IPO investors on the first trading day is 1.3\%, and when the target is announced, SPAC investors receive an average return of 3.74\%.

{\bf Money Left on the Table.}
We construct a measure \emph{money left on the table} (``MLOT''), which is the direct analog of the underpricing measure used in the traditional IPO literature.  Specifically, we MLOT denotes the difference between the closing-day price on the first trading day\footnote{As in the IPO literature, the motive for using first-day closing prices is weak-form efficiency, i.e., the first-day closing price reflects the best estimate of future cash flows using all available information. Longer holding period horizons would capture the same phenomena while introducing more noise into our tests.} of the post-merger company and the offer price presented to PIPE investors by the SPAC, and this difference is multiplied by the number of securities held by the PIPE investors. This calculation considers both common shares and warrants acquired by the PIPE investors. To determine the rate of underpricing, we normalize the MLOT by PIPE investors' total investment amount.\footnote{Although the contracts with PIPE investors are signed before the merger is closed, they represent a pre-commitment price, implying that the PIPE investors deploy their capital only at the merger's closure. This arrangement mirrors a "one-day" return that traditional IPO investors experience.} We carry out the MLOT and underpricing calculations for all three forms of PIPE investors (pre, at, and post).

\begin{table}[!htbp]
    \centering
    \caption{Summary Statistics}
    \centering
    \scalebox{0.90}{
\begin{tabular}{l*{1}{cccccc}}
\toprule
                    &        Mean&          Std.Dev.&         P10&         P50&         P90&       Obs.\\
\midrule
\\[-1.8ex]
\textbf{SPAC Measures} \\
IPO proceeds (mm USD)&      272.42&      218.22&      100.00&      230.00&      440.00&       1,072\\
No. of Warrant      &        0.44&        0.24&        0.20&        0.50&        0.75&       1,072\\
No. of Right        &        0.01&        0.03&        0.00&        0.00&        0.00&       1,072\\
Overallotment (\%)  &       11.59&        5.37&        0.00&       15.00&       15.00&       1,072\\
Listing-day return (\%)&        1.30&        3.02&       -0.70&        0.40&        4.00&         857\\
Days searching      &      361.71&      215.79&      114.00&      329.50&      677.00&         614\\
Days remaining      &      340.98&      189.87&      120.00&      303.00&      611.00&         614\\
Announcement-day return (\%)&        3.74&       12.56&       -0.68&        0.34&        9.95&         586\\
Redeemption(\%)     &       56.54&       37.49&        0.00&       70.00&       97.00&         445\\
Days deSPAC         &      167.99&       67.06&      100.00&      160.00&      241.00&         474\\
\\[-1.8ex]
\textbf{PIPE Participation} \\
$\mathbbm{1}$\{PIPE (pre)\}&        0.10&        0.30&        0.00&        0.00&        0.00&         473\\
$\mathbbm{1}$\{PIPE (at)\}&        0.66&        0.47&        0.00&        1.00&        1.00&         473\\
$\mathbbm{1}$\{PIPE (post)\}&        0.12&        0.33&        0.00&        0.00&        1.00&         473\\
$\mathbbm{1}$\{Premium PIPE (at)\}&        0.50&        0.50&        0.00&        1.00&        1.00&         473\\
$\mathbbm{1}$\{Non-Premium PIPE (at)\}&        0.65&        0.48&        0.00&        1.00&        1.00&         473\\
No. of PIPE (at)    &       10.76&       11.83&        0.00&        8.00&       26.00&         472\\
No. of Premium PIPE (at)&        3.02&        4.36&        0.00&        1.00&       10.00&         472\\
No. of Non-Premium PIPE (at)&        7.75&        8.64&        0.00&        6.00&       18.00&         472\\
\\[-1.8ex]
\textbf{PIPE Investment/IPO Proceeds} \\
PIPE (pre) (\%)     &        2.93&       11.22&        0.00&        0.00&        0.00&         473\\
Premium PIPE (at) (\%)&       17.84&       32.85&        0.00&        0.00&       52.39&         473\\
Non-Premium PIPE (at) (\%)&       41.94&       58.64&        0.00&       28.57&      101.16&         473\\
PIPE (post) (\%)           &     7.45       &    30.39        &        0.00&        0.00&       22.68&         473\\
\\[-1.8ex]
\textbf{MLOT (mm USD)} \\
Public Investors           &      137.35&      323.71&       -2.96&       33.85&      364.26&         162\\
Sponsors          &      109.76&      161.45&       23.00&       77.30&      210.74&         160\\
PIPE          &      201.22&    1,200.54&      -25.60&        3.75&      248.17&         358\\
PIPE (pre)         &       29.37&      120.76&      -10.52&        3.17&       35.70&          45\\
PIPE (at)         &      193.07&    1,135.58&      -25.36&        4.77&      248.17&         303\\
PIPE (post)$^*$        &        8.07       &   42.62          &      -15.74&        0.25&       31.76&          56\\
Premium PIPE (at)         &       76.63&      374.15&      -10.22&        1.68&      129.46&         231\\
Non-Premium PIPE (at)       &      134.65&      838.67&      -17.73&        3.40&      138.48&         303\\
\\[-1.8ex]
\textbf{MLOT/Investment (\%)} \\
PIPE        &       91.26&      472.71&      -16.50&        4.47&       85.80&         358\\
PIPE (pre)        &       70.33&      289.37&       -8.05&        4.36&       91.80&          45\\
PIPE (at)         &       63.74&      307.47&      -16.00&        4.20&       80.90&         303\\
PIPE (post)$^*$         &     70.74          &  285.16         &      -36.02&        0.80&      110.60&          56\\
Premium PIPE (at)       &       64.05&      291.75&      -13.00&        5.30&       94.90&         225\\
Non-Premium PIPE (at)      &       64.10&      308.46&      -16.00&        4.20&       80.90&         301\\
\bottomrule
\end{tabular}}
\begin{figurenotes}
    The sample includes 1,072 SPACs that went through the IPO process between January 2010 and December 2021. All variables are defined in Table \ref{tab:def}. $(^*)$ We exclude a PIPE (post) investor named \emph{Quinpario Acquisition Corporation 2}, with  unusually large returns.
    \end{figurenotes}\label{tab:sum_stat}
\end{table}

Table \ref{tab:sum_stat} highlights the MLOT for all three types of PIPE investors. 
Investing in a SPAC generates an average of 
\$201.22 million for PIPE investors. Out of the 45 SPAC deals in which they participate, PIPE (pre) investors make \$29.37 million on average. From the 303 SPAC deals in which they take part, PIPE (at) investors earn \$193.07 million per deal. PIPE (post) investors participate in 56 deals and earn \$8.07 million per deal on average.

A SPAC deal generates 
an average return of 91.26\%, where PIPE (pre) investors earn 70.33\%, PIPE (at) investors earn 63.74\%, and PIPE (post) investors earn 70.74\%.

{\bf Premium PIPE Investors.} 
In total, there are 2,005 PIPE investors in our sample. Out of them, 76 invested in PIPE (pre) transaction, 249 in PIPE (post), and 1,826 invested in at least one PIPE (at) transaction, but only a few large institutional investors participated frequently and invested significantly more than others. 
To capture this distinction, we use k-means clustering on the total investments in all SPACs and the number of SPACs to classify PIPE investors into two groups. We refer to a subset of PIPE investors who frequently participate as ``premium'' PIPE investors and the rest as non-premium PIPE investors. There are 34 premium PIPE investors in our sample (Table \ref{tab:premium_investor}).

\begin{table}[!t]
    \centering
    \caption{Premium PIPE Investors}
    \centering
    \scalebox{0.9}{
\begin{tabular}{lcc}
\toprule
                                                        Investor &  \#SPAC &  Volume (million USD) \\
\midrule
                         Fidelity Management \& Research Company &      69 &        4,342 \\
                                                BlackRock, Inc. &      57 &        2,317 \\
                        Capital Research and Management Company &      29 &        2,231 \\
                                  Alyeska Investment Group L.P. &     106 &        1,704 \\
                                                 Morgan Stanley &      14 &        1,238 \\
                                     Millennium Management, LLC &      77 &        1,059 \\
                                            Luxor Capital Group &      52 &          972 \\
              Blackstone Inc. (f/k/a The Blackstone Group Inc.) &      34 &          965 \\
                                           Koch Industries Inc. &      21 &          926 \\
                                               Third Point, LLC &      12 &          910 \\
                                    Hedosophia Services Limited &      23 &          910 \\
                                      T. Rowe Price Group, Inc. &      27 &          899 \\
                                           SoftBank Group Corp. &      16 &          795 \\
Franklin Resources, Inc. (d/b/a Franklin Templeton Investments) &      24 &          741 \\
                                    Senator Investment Group LP &      36 &          713 \\
                                 Moore Capital Management, Inc. &      66 &          629 \\
                                     Neuberger Berman Group LLC &      20 &          626 \\
                                      Baron Capital Group, Inc. &      25 &          621 \\
                                                    Citadel LLC &      51 &          604 \\
                               Hudson Bay Capital Management LP &      56 &          577 \\
                               Suvretta Capital Management, LLC &      31 &          576 \\
                                                         UBS AG &      57 &          494 \\
                             Wellington Management Company, LLP &      26 &          437 \\
                               Heights Capital Management, Inc. &      65 &          431 \\
                                           Magnetar Capital LLC &      39 &          424 \\
                                          D.E. Shaw \& Co., L.P. &      38 &          411 \\
                             Monashee Investment Management LLC &      61 &          408 \\
                                Ghisallo Capital Management LLC &      62 &          389 \\
                                         Jane Street Group, LLC &      49 &          368 \\
                                Park West Asset Management, LLC &      27 &          342 \\
                                               Kepos Capital LP &      51 &          286 \\
                                             Linden Advisors LP &      43 &          282 \\
                               Schonfeld Strategic Advisors LLC &      32 &          224 \\
                              BlueCrest Capital Management Ltd. &      31 &          181 \\
                              \bottomrule
\end{tabular}
}
\begin{figurenotes}  
        This table lists the name of premium PIPE investors identified using k-means clustering based on the number of SPAC deals and total investment volumes using data on SPACs that went through an IPO from January 2010 to December 2021. 
\end{figurenotes}
\label{tab:premium_investor}
\end{table}

Table \ref{tab:sum_stat} offers summary statistics for premium and non-premium PIPE investors. A standard SPAC deal usually involves about ten PIPE (at) investors, typically divided into three premium and seven non-premium investors. In the 231 SPAC transactions in which they participate, premium investors realize an average MLOT of \$76.63 million with an average return of 64.05\%.

In contrast, non-premium investors participate in 303 SPAC deals, generating an average MLOT of \$134.65 million and achieving an average return of 64.10\%. The outcomes for non-premium investors display greater variability in terms of dollar values.

It is important to note that within each SPAC, both premium and non-premium investors are given the same pricing terms, indicating that the difference in returns within each SPAC arises purely from the variance in their investment amounts.

This comparison is consistent with the quid pro quo arrangement where non-premium investors attain high returns on successful deals, offset by low returns on less successful ones. In contrast, premium investors see more steady returns due to their superior information. 

\section{Empirical Results\label{section:results}}
In this section, we investigate whether the data are consistent
with the hypotheses stated in Section \ref{section:hypotheses}. 
We begin measuring underpricing at the SPAC IPO level and compare it with the traditional IPO underpricing. 
Then, we show that the premium PIPE investors who participate in business combinations generate value-relevant information, whereas non-premium investors engage in a quid pro quo relationship. 
We end by quantifying the benefit of the quid pro quo relationship in terms of lowering the liquidation risk of future SPACs. 

\subsection{Underpricing}
\subsubsection{Underpricing during SPAC IPO}

We first examine underpricing during the SPAC IPO. Table \ref{tab:sum_stat} reports that the average
equal-weighted first-day return of SPACs is 1.3\%. Although our data start from 2010,
most SPAC IPOs occurred between 2020 and 2021. In 2020, 248 SPACs went public, with an
average equal-weighted first-day return of 1.3\%; in 2021, 613 SPACs went public, and their
average equal-weighted first-day return was 1.51\%. Although these measures are statistically
different from zero (p-values $\approx0$ in both cases), they are small.\footnote{The average equal-weighted first-day return for traditional IPOs was 41.6\% in 2020 and 32\% in 2021; see \url{https://site.warrington.ufl.edu/ritter/files/IPOs-Underpricing.pdf} (accessed 09/22/2022).}
In addition, SPACs almost always trade at \$10 per share until the target announcement date. This uniformity is inconsistent with any information about sponsor quality revealed during the SPAC IPO.

\subsubsection{Underpricing during de-SPAC}

This section focuses on the underpricing associated with the de-SPAC stage. 
We say that a private company's shares are underpriced if the first-day trading value of the post-merger company's securities held by all SPAC investors (including sponsors, PIPEs, and original SPAC investors who did not redeem) exceeds the proceeds invested in the private firm.  

There are at least two sources of asymmetric information at this stage. 
First, the target company may be better informed about its business prospects than anyone else \citep{smith1986investment, booth1986capital}. Second, the PIPE investors may also have better information about factors affecting the target firm than the public investors. In response to PIPE investors' information, sponsors may underprice the securities issued to those investors as an incentive payment for their value-relevant information. We focus on the second type of information.

From Table \ref{tab:sum_stat}, we see that, on average, a private company that uses the SPAC route to access the public market leaves \$126.93  million MLOT to the sponsors and \$201.22 million to the PIPE investors. 
Furthermore, we also see that PIPE investors who participate in different stages of SPAC are compensated differently. During the SPAC IPO, PIPE (pre) investors earn an average of \$29.37 million through forward-purchase agreements, but PIPE (at) investors get, on average, a total of \$193.07 million per deal. 
After the target is disclosed, PIPE (post) investors who participate receive an additional \$8.07 million MLOT.  
Even among PIPE (at) investors, approximately one-third of the MLOT accrues to 34 premium investors. 

Overall, the SPAC shares offered to PIPE (at) investors, especially the premium PIPE investors, during the de-SPAC phase are underpriced. These data features are consistent with PIPE (at) investors participating in the de-SPAC stage being privately informed about the quality of the target company. 

\subsection{Certification and Agency Theories}
\subsubsection{Redemption Rates}

In this section, we examine Hypothesis \ref{hyp:info_ipo}-1 and Hypothesis \ref{hyp:agency}-1 using the following regression:
\begin{eqnarray}
\text{Redemption rate}_i&=&\alpha_0+\alpha_1\times\% \text{Premium PIPE (at)}_i+\alpha_2\times\%\text{Non-Premium PIPE (at)}_i\nonumber \\
&&+\alpha_3\times\%\text{PIPE (pre)}_i+\alpha_4\times\%\text{PIPE (post)}_i+ {\bf X}^{\top}\alpha_5+\varepsilon_i, \label{eq:redemption_OLS}
\end{eqnarray}
where $\text{Non-Premium PIPE (at)}_i$ denotes the investment allocated to premium PIPE (at) investors as a fraction of the total IPO proceeds in SPAC $i$, similarly for other allocations, and ${\bf X}$ is a vector of covariates including total IPO proceeds and fixed effects that may vary across specifications. 
For example, if premium investors generate value-relevant information, the higher the share allocated to premium investors, the lower the redemption rate, i.e., $\alpha_1 < 0$. For inference, we cluster the robust standard errors at the sponsor level to capture correlations within a sponsor but across investors and SPACs.

Table \ref{tab:t_redemption_premium} provides evidence consistent with both hypotheses. In column (1), we include only the allocations to four types of PIPE investors. In column (2), we control for the size of the SPAC IPO and the liquidation risk of the SPAC when the target is announced using Days Remaining until the expiration. In column (3), we control for the quarter-fixed effects. In column (4), we control for the target company's sector fixed effects to alleviate the problem that investors may have varying investment preferences across industries.

\begin{table}[!t]
\centering
\caption{Redemption Rates}
\centering
\scalebox{1}{
\begin{tabular}{l*{5}{c}}
\toprule
                    &\multicolumn{1}{c}{(1)}&\multicolumn{1}{c}{(2)}&\multicolumn{1}{c}{(3)}&\multicolumn{1}{c}{(4)}&\multicolumn{1}{c}{(5)}\\
\hline
PIPE (pre) (\%)     &       0.125         &       0.192         &       0.148         &       0.156\sym{*}  &       0.148         \\
                    &     (0.123)         &     (0.119)         &     (0.116)         &     (0.095)         &     (0.094)         \\
Premium PIPE (at) (\%)&      -0.419\sym{***}&      -0.380\sym{***}&      -0.366\sym{***}&      -0.214\sym{***}&      -0.219\sym{***}\\
                    &     (0.069)         &     (0.064)         &     (0.059)         &     (0.050)         &     (0.051)         \\
Non-Premium PIPE (at) (\%)&      -0.013         &      -0.009         &      -0.002         &      -0.037         &      -0.036         \\
                    &     (0.034)         &     (0.035)         &     (0.034)         &     (0.030)         &     (0.031)         \\
PIPE (post) (\%)    &       0.120\sym{***}&       0.103\sym{***}&       0.114\sym{***}&       0.148\sym{***}&       0.148\sym{***}\\
                    &     (0.035)         &     (0.031)         &     (0.033)         &     (0.035)         &     (0.035)         \\
\hline
Days Remaining & \xmark &\checkmark &\checkmark&\checkmark&\checkmark\\
IPO Month FEs          & \xmark & \xmark  & \checkmark  & \checkmark &\checkmark\\
Merger Sector FEs    & \xmark &\xmark   & \xmark  & \checkmark &\checkmark\\
Assortative Matching & \xmark &\xmark   & \xmark  & \xmark &\checkmark\\
\hline
Adj. $ R^2 $        &       0.161         &       0.182         &       0.183         &       0.400         &       0.398         \\
Obs.                &         444         &         444         &         444         &         430         &         430         \\
\bottomrule
\end{tabular}}%
\label{tab:t_redemption_premium}
\begin{figurenotes}
This table shows the results from estimating   (\ref{eq:redemption_OLS}) using SPACs that went through an IPO from January 2010 to December 2021. The independent variable is the investment amount as a percentage of the SPAC's IPO proceeds for all PIPE (pre) investors, premium PIPE (at) investors, non-premium PIPE (at) investors, and all PIPE (post) investors.
Premium PIPE (at) investors are listed in Table \ref{tab:premium_investor}.  
All other variables are defined in Table \ref{tab:def}. 
All unbounded continuous variables are winsorized at their  $1^{st}$ and $99^{th}$ percentile values.
Robust standard errors are clustered at the sponsor level and are reported in parentheses. In column (5), we use 1,000 Bootstrap samples to estimate the standard errors.. $^*, ^{**}$, and $^{***}$ denote \textit{p}-values less than 0.1, 0.05, and 0.01, respectively.
\end{figurenotes}
\end{table}

We find that the allocations to premium PIPE (at) investors negatively correlate with the redemption rates, and this result is robust across all specifications. In particular, the estimates suggest that one standard deviation (32.85 pps) increase in premium PIPE investors' participation is associated with a 7.03 pps decrease in the redemption rate, which equals 12.4\% of the sample mean.  This evidence is consistent with both Hypothesis \ref{hyp:info_ipo}-1 and Hypothesis \ref{hyp:agency}-1.

A threat to identification is that the sample of sponsor and premium PIPE investors is not random. If better sponsors work with better investors and allocate more to those investors, the estimate of $\alpha_1$ will suggest a larger (negative) decrease in the redemption rate. In other words, the estimate $-0.214$ in column (4) may also capture the effect of matching where certain sponsors consistently secure favorable deals, which they then allocate to a select group of premium investors. 
To address this issue, we estimate the fixed effect of each sponsor and premium investor following the approach of \cite{abowd1999high} and include those fixed effects as additional controls in (\ref{eq:redemption_OLS}). 

In particular, we estimate the following two-way fixed effect model of 
\begin{eqnarray}
\text{MLOT}_{p, s,t}=\mathbf{Q}_{p,s,t}^\top\mathfrak{b}+\texttt{fe}_p+\texttt{fe}_{s}+\varrho_{p,s,t} \label{eq:akm}
\end{eqnarray}
that relates the money left on the table that investor $p$ gets from working with sponsor $s$ in time $t$, $\text{MLOT}_{p,s,t}$, with their fixed-effects, $\texttt{fe}_p$ and $\texttt{fe}_{s}$, respectively, and observed characteristics $\mathbf{Q}$ that includes quarter fixed effects, target sector fixed effects, the SPAC size, and the number of PIPE investors.
Heuristically speaking, we use the returns, $\text{MLOT}_{p,s,t}$, to disentangle how much of it is attributable to the sponsor and investor. Most sponsors initiate more than one SPAC and work with different sets of investors each time. Likewise, investors work with different sponsors. These switching patterns identify these fixed effects. 

A sponsor works with several premium investors, so to control for the quality of the sponsor and premium investors, for each SPAC, we use the sponsor fixed effect and the average fixed effect among all premium investors as additional regressors that capture the sponsor and premium investor qualities. 
Consistent with the matching hypothesis, these two fixed effects are positively correlated (regression coefficient $1.2$, p-value $<0.01$). In contrast, the sponsor fixed effect and the average fixed effect among all non-premium investors are negatively correlated (regression coefficient $-0.49$, p-value $<0.01$).

The estimates with the additional controls are in Table \ref{tab:t_redemption_premium} column (5). 
Comparing the estimates in columns (4) and (5), we can see that our result is qualitatively similar. The estimates suggest that one standard deviation (32.85 pps) increase in premium PIPE (at) investors' participation is associated with a 7.19 pps decrease in the redemption rate, which equals 12.7\% of the sample mean.  This evidence is consistent with both Hypothesis \ref{hyp:info_ipo}-1 and Hypothesis \ref{hyp:agency}-1.

The other coefficients in Table \ref{tab:t_redemption_premium} are similarly intuitive.  
We also find no association between non-premium PIPE (at) or PIPE (pre) allocations and redemption rates.   Thus, non-premium investors act neither in a certification role nor an agency role.  Additionally, PIPE (pre) investors cannot certify an issue, as there is no target identified at that point.  Finally, PIPE (post) loads positively in all specifications, reflecting instances in which PIPE investors are brought in to shore up funding shortfalls caused by redemption.

\subsubsection{Announcement-day Return}

To test the second part of Hypothesis \ref{hyp:info_ipo}-1 about the announcement-day return, we use the same model as (\ref{eq:redemption_OLS}) except the dependent variable is $\text{Announcement-day return}_i$ and the control variable $\%\text{PIPE (post)}_i$ is excluded.  
The estimation results are shown in Table \ref{tab:t_r_ann_premium}. 
There is a strong positive association between announcement-day returns (\%) and allocations to premium PIPE (at) investors. In particular, the estimates from our preferred specification in Table \ref{tab:t_r_ann_premium} column (5) suggest that one standard deviation (32.85 pps) increase in premium PIPE (at) investors' participation is associated with a 4.14 pps increase in the SPAC's announcement-day return, (110.7\% of the sample mean). Thus, the effect of premium investors on announcement-day return is economically and statistically significant.

\begin{table}[!t]
\centering
\caption{Announcement-day Return}
\centering
\scalebox{1}{
\begin{tabular}{l*{5}{c}}
\toprule
                    &\multicolumn{1}{c}{(1)}&\multicolumn{1}{c}{(2)}&\multicolumn{1}{c}{(3)}&\multicolumn{1}{c}{(4)}&\multicolumn{1}{c}{(5)}\\
\hline
PIPE (pre) (\%)     &      -0.118\sym{***}&      -0.119\sym{***}&      -0.099\sym{***}&      -0.072         &      -0.068         \\
                    &     (0.042)         &     (0.042)         &     (0.037)         &     (0.044)         &     (0.043)         \\
Premium PIPE (at) (\%)&       0.148\sym{***}&       0.140\sym{***}&       0.136\sym{***}&       0.124\sym{***}&       0.126\sym{***}\\
                    &     (0.041)         &     (0.041)         &     (0.038)         &     (0.040)         &     (0.040)         \\
Non-Premium PIPE (at) (\%)&      -0.003         &      -0.007         &      -0.010         &      -0.002         &      -0.002         \\
\midrule
Log IPO Proceeds & \xmark &\checkmark &\checkmark&\checkmark&\checkmark\\
Days Remaining & \xmark &\checkmark &\checkmark&\checkmark&\checkmark\\
IPO Month FEs          & \xmark & \xmark  & \checkmark  & \checkmark &\checkmark\\
Merger Sector FEs    & \xmark &\xmark   & \xmark  & \checkmark &\checkmark\\
Assortative Matching & \xmark &\xmark   & \xmark  & \xmark &\checkmark\\
\midrule
Adj. $ R^2 $        &       0.099         &       0.104         &       0.098         &       0.095         &       0.091         \\
Obs.                &         457         &         457         &         431         &         417         &         417         \\
\bottomrule
\end{tabular}}%
\label{tab:t_r_ann_premium}
\begin{figurenotes}
This table shows the results from estimating (\ref{eq:redemption_OLS}), announcement-day return as the dependent variable, using SPACs that went through an IPO from January 2010 to December 2021. The key independent variable is the investment amount as a percentage of the SPAC's IPO proceeds for all PIPE (pre) investors, premium PIPE (at) investors, non-premium PIPE (at) investors, and all PIPE (post) investors.
Premium PIPE (at) investors are listed in Table \ref{tab:premium_investor}.  
All other variables are defined in Table \ref{tab:def}. 
All unbounded continuous variables are winsorized at their  $1^{st}$ and $99^{th}$ percentile values.
Robust standard errors are clustered at the sponsor level and are reported in parentheses. In column (5), we use 1,000 Bootstrap samples to estimate the standard errors. $^*, ^{**}$, and $^{***}$ denote \textit{p}-values less than 0.1, 0.05, and 0.01, respectively.
\end{figurenotes}
\end{table}

Furthermore, the announcement-day returns are uncorrelated with the allocations to non-premium PIPE (at) investors. 
Again, this finding is consistent with the hypothesis that larger investment by premium PIPE (at) investors acts as a certification of the quality of the deal. In contrast, investment by non-premium PIPE (at) is non-informative. 

We also observe a negative correlation between announcement-day returns and PIPE (pre) investors' allocations. 
Specifically, the estimates in column (5) suggest that one standard deviation (11.22 pps) increase in PIPE (pre) investors' participation is associated with 0.84 pps lower in announcement-day return, which equals 22.5\% of the sample mean.

Likewise, the hypothesis predicts a negative market reaction to the announcement of PIPE (post) investments. We use an event study approach to assess this claim. 
In particular, we estimate the cumulative abnormal returns (CAR) \citep{CampbellLoMacKinlay1997} associated with announcements about PIPE investors while also announcing the target. 
The estimation period is from the SPAC's unit separation date to the event window.

Table \ref{tab:eventstudy} reports the average of the CAR and their respective standard deviation across all SPACs using a three-day, five-day, and 10-day window around the event date. For each event window, we estimate three different asset-pricing models: CAPM, Fama-French three-factor, and Fama-French five-factor models. The average CAR estimates are not statistically different from zero in all cases. 
This result suggests that the investors do not respond to the announcement of PIPE (post), likely because almost all of them are non-premium and unlikely to add information.

\begin{table}[t!]
\caption{Cumulative (average) Abnormal Returns}
\medskip
\raggedright
\centering
\scalebox{1}{
\begin{tabular}{lccc}
\toprule
Event Window&\multicolumn{1}{c}{CAPM}&\multicolumn{1}{c}{FF3}&\multicolumn{1}{c}{FF5}\\
\midrule
{[}-3, 3{]}   &0.71	    	&0.74		 &0.75    \\
			  &(0.80)		&(0.79)		 &(0.78)  \\
{[}-5, 5{]}   &-1.04	    &-1.04		 &-1.06   \\
			  &(1.14)		&(1.13)		 &(1.12)  \\
{[}-10, 10{]} &-2.37	    &-2.30		 &-2.28   \\
			  &(1.92)		&(1.92)		 &(1.91)  \\			
\bottomrule
\end{tabular}}
\begin{figurenotes}
This table reports the SPAC share's cumulative average abnormal return in percentage for three different event windows around announcements of 81 PIPE (post) transactions. Here, CAPM stands for capital asset pricing model, and  FF3 and FF5 stand for Fama-French three- and five-factor models, respectively. The cross-sectional standard deviations of the cumulative abnormal returns are reported in parentheses. $^*, ^{**}$, and $^{***}$ denote \textit{p}-values less than 0.1, 0.05, and 0.01, respectively.

\end{figurenotes}
\label{tab:eventstudy}
\end{table}


\subsection{Insurance Theory}
Next, we test Hypothesis \ref{hyp:insurance} that there is a quid pro quo arrangement between sponsors and relationship investors. In particular, if the sponsor allocates securities to certain investors as a favor, and if the past MLOT is high, the sponsor will likely ask this investor to finance a weak deal today. 
We verify that the probability that a sponsor allocates securities to such a relationship investor increases if the investor has higher MLOT from the past deals \emph{and} the SPAC deal is cold. Lastly, we show these relationships reduce the SPAC liquidation risks.

\subsubsection{Mean-reverting MLOT}
As a first step, we assess if the MLOT exhibits mean reversion for the non-premium PIPE (at) investors, which is a testable implication of quid pro quo. In contrast, under the information production hypothesis, the correlation between the current MLOT and past MLOT should be zero or positive if there is an information spillover \citep{BenvenisteLjungqvistWilhelmYu2003}.  

However, we need repeated deals between a sponsor and a PIPE investor to test the implication. In Table \ref{tab:t_sponsor}, we list the top 20 sponsors in our sample based on the total number of initialized and completed SPAC deals. In our sample, 139 sponsors have initialized more than one SPAC, totaling 439 SPACs.
Furthermore, non-premium PIPE investors are more likely to participate in repeat deals (with the same sponsor) than premium investors. 

To gauge the difference, consider the following.
On average, the probability that a non-premium investor from the previous deal will participate in the current deal is 17\%. 
There are around eight non-premium investors in a SPAC, so one of these eight non-premium investors in the current deal is from the previous one. If these investors were randomly (uniformly) participating in a SPAC deal, the probability would be much smaller because there are 1,792 non-premium investors in our sample. If non-premium investors participate randomly, the probability that at least one non-premium investor participated previously is negligible.

\begin{table}[t!]
     \centering
     \caption{Top 20 Sponsors}
     \medskip
     \raggedright
     \centering
     \scalebox{1}{
\begin{tabular}{lclc}
\toprule
\multicolumn{2}{c}{Initialized} &\multicolumn{2}{c}{Completed} \\
\cmidrule(lr){1-2} \cmidrule(lr){3-4}
                 Sponsor &  \#SPAC &          Sponsor &  \#SPAC \\
\midrule
            The Gores Group &      13 &             The Gores Group &              9 \\
            Cohen \& Company &      11 &                     Chardan &              7 \\
             Social Capital &      10 &              Social Capital &              6 \\
     Hennessy Capital Group &       8 &             Cohen \& Company &              6 \\
          Cantor Fitzgerald &       8 &                         TPG &              5 \\
                        TPG &       8 &           Cantor Fitzgerald &              5 \\
              Michael Klein &       8 & Riverstone Investment Group &              5 \\
        Jonathan J. Ledecky &       7 &      Hennessy Capital Group &              5 \\
                    Chardan &       7 &                Harry L. You &              5 \\
                 Hedosophia &       6 &       Eagle Equity Partners &              5 \\
               Harry L. You &       6 &               Michael Klein &              4 \\
Riverstone Investment Group &       6 &        Jonathan S. Huberman &              4 \\
  Fortress Investment Group &       6 &                  Hedosophia &              4 \\
          GigCapital Global &       6 &           GigCapital Global &              4 \\
   Apollo Global Management &       6 &                   Chinh Chu &              4 \\
                 Bill Foley &       6 &             Niccolo de Masi &              4 \\
                  Chinh Chu &       5 &                  Bill Foley &              4 \\
       Jaws Estates Capital &       5 &    Apollo Global Management &              4 \\
 Craig-Hallum Capital Group &       5 &     Union Acquisition Group &              3 \\
            Niccolo de Masi &       5 &              Casdin Capital &              3 \\
\bottomrule
\end{tabular}\label{tab:t_sponsor}}

\begin{figurenotes}
          This table presents the top 20 sponsors in terms of the total number of SPACs initialized and completed.
\end{figurenotes}
\end{table}

The probability gap reverses for the premium investors. The probability that a premium investor from the previous deal will participate in the current deal is 36.60\%. There are, on average, three premium investors in a SPAC, so a sponsor includes one premium investor from the previous deal. There are 34 premium investors in our sample, so if they participate randomly in a SPAC, then the probability that at least one premium investor in the current deal has participated in the previous deal by the same sponsor is 90.31\%.

Let $\mu$ be the average MLOT for all PIPE (at) investors in our sample. 
Table \ref{tab:t_sum_sponsor_pipe} is the summary statistics. On average, 28\% of SPACs have at least one premium PIPE investor. 
Then, we estimate the following linear model,
\begin{eqnarray}
\text{MLOT}_{p, s, i}&=&\theta_0+\theta_1\times (\text{lag-MLOT}_{p, s, i}-\mu)+{\bf X}^{\top} \theta_2+\eta_{p, s, i},\label{eq:lagmlot1}
\end{eqnarray}
where \text{MLOT}$_{p, s, i}$ is the MLOT that PIPE (at) investor $p$ gets in the current SPAC deal $i$ initialized by the sponsor $s$, and  $\text{lag-MLOT}_{p, s, i}$ is the MLOT from the most recent deal in the past. As before, ${\bf X}$ includes several covariates, including fixed effects (see Table \ref{tab:mlot_last}). 

\begin{table}[t!]
    \centering
    \caption{Summary Statistics of PIPE Investors}
    \centering
    \scalebox{1}{
\begin{tabular}{l*{1}{cccccc}}
\toprule
                    &        Mean&          Std.Dev.&         P10&         P50&         P90&       Obs.\\
\midrule
At Least 1 Premium    &        0.28&        0.45&        0.00&        0.00&        1.00&       5,812\\
\\[-1.8ex]
\textbf{All PIPE Investors} \\
Investment Amount (\$mm)   &       16.67&       37.62&        1.00&        7.00&       35.00&       5,535\\
MLOT (\$mm)                &        9.76&       75.33&       -0.98&        0.28&       13.05&       4,816\\
No. of Past SPACs      &        0.13&        0.47&        0.00&        0.00&        0.00&       5,812\\
MLOT (past)          &        2.48&       23.66&        0.00&        0.00&        0.00&       5,812\\
\\[-1.8ex]
\textbf{Premium PIPE Investors} \\
Investment Amount (\$mm)   &       20.50&       38.79&        3.00&       10.00&       45.00&       1,609\\
MLOT (\$mm)                &       11.01&       45.60&       -1.17&        0.58&       23.97&       1,462\\
No. of Past SPACs       &        0.20&        0.53&        0.00&        0.00&        1.00&       1,647\\
MLOT (past)          &        3.98&       19.82&        0.00&        0.00&        2.55&       1,647\\
\\[-1.8ex]
\textbf{Non-premium PIPE Investors} \\
Investment Amount (\$mm)   &       15.10&       37.01&        0.75&        5.00&       30.00&       3,926\\
MLOT (\$mm)                &        9.22&       85.10&       -0.96&        0.17&       10.37&       3,354\\
No. of Past SPACs       &        0.10&        0.43&        0.00&        0.00&        0.00&       4,165\\
MLOT(past)          &        1.88&       25.00&        0.00&        0.00&        0.00&       4,165\\
\bottomrule
\end{tabular}\label{tab:t_sum_sponsor_pipe}}
\begin{figurenotes}
    This table contains summary statistics for SPACs that went through an IPO from January 2010 to December 2021. 
    Premium PIPE investors are identified using k-means clustering and listed in Table \ref{tab:premium_investor}.
    All variables are defined in Table \ref{tab:def}.
    \end{figurenotes}

\end{table}

According to the certification hypothesis and agency theory, the deviation of lagged MLOT from its mean is either uncorrelated with the current MLOT; hence $\theta_1 =0$ in \eqref{eq:lagmlot1}. Under the insurance hypothesis, however, the MLOT should increase if the past deal was lower than the average and vice versa; hence, $\theta_1<0$. 
The results are in column (1) of Table \ref{tab:mlot_last}. Consistent with the insurance hypothesis, MLOT exhibits mean reversion.

\begin{table}[t!]
\centering
\caption{Mean-Reverting MLOT}
\medskip
\raggedright

\medskip
\centering
\scalebox{1}{
\begin{tabular}{l*{3}{c}}
\toprule
                    &\multicolumn{3}{c}{MLOT}\\
                    &\multicolumn{1}{c}{(1)}&\multicolumn{1}{c}{(2)}&\multicolumn{1}{c}{(3)}\\
\hline
All PIPE Investors    &      -0.099\sym{*}&                     &                \\
                       &     (0.052)               &                     &               \\
Premium PIPE Investors &                     &       -0.025          &   0.235\sym{**}                      \\
                    &                     &     (0.062)        &          (0.109)          \\
Non-premium PIPE Investors &                     &     -0.218\sym{***}& -2.729\sym{***}         \\
                    &                     &     (0.076)          &   (0.756)                    \\
Instrumental Variables & \xmark & \xmark & \checkmark  \\
\midrule
F-stat (first-stage)&  &  & 92.82 \\
Adjusted $ R^2 $        &       0.418         &       0.419        &       
\\
Observations                &       3,790         &       3,790         &       3,790         \\
\midrule
&& Mean &Std.Dev. \\
\midrule
$\text{Non-Premium}\times$(lag-MLOT-$\mu$)&&-8.12  &  5.41\\
(lag-MLOT-$\mu$)$\times$(lag$^2$-MLOT-$\mu$)& &121.96  &  85.54\\
(lag$^2$-MLOT-$\mu$)$^2$ &&135.78 &   5.77\\
(lag-MLOT-$\mu$)$^2$ &&166.93 &   297.54\\
\bottomrule
\end{tabular}}%
\label{tab:mlot_last}
\begin{figurenotes}
This table shows results from estimating (\ref{eq:lagmlot2}) using SPACs that went through IPO between January 2010 to December 2021. Each observation is at (SPAC, sponsor, PIPE$_{at}$ investor) level. The dependent variable, MLOT, measures the money left on the table the PIPE$_{at}$ investor makes from the SPAC deal. The key independent variables are the MLOT earned with the same sponsor in the last deal for all PIPE$_{at}$ investors, premium PIPE$_{at}$ investors, and non-premium PIPE$_{at}$ investors. 
Additional controls include the log of IPO proceeds, the number of PIPE investors, the number of premium PIPE investors, days remaining, and separate fixed effects for the month, sponsor, and PIPE.  
All continuous variables are winsorized at the $1^{st}$ and $99^{th}$ percentile values. All other variables are defined in Table \ref{tab:def}.  
Robust standard errors are reported in parentheses. $^*, ^{**}$, and $^{***}$ denote \textit{p}-values less than 0.1, 0.05, and 0.01, respectively.
\end{figurenotes}
\end{table}

Next, we differentiate between premium and non-premium investors and estimate 
\begin{eqnarray}
\text{MLOT}_{p, s, i}
&=&\delta_0+\delta_1\times (\text{lag-MLOT}_{p, s, i}-\mu)\notag\\&&+\delta_2\times \text{Non-Premium}_p\times(\text{lag-MLOT}_{p, s, i}-\mu)+{\bf X}^{\top}\delta_3+\tilde{\eta}_{p, s, i},\label{eq:lagmlot2}
\end{eqnarray}
where  $\text{Non-Premium}_p$ is a binary variable equal to one if the PIPE (at) investor $p$ is a non-premium investor. The results from the estimation are in column (2) of Table \ref{tab:mlot_last}.
Comparing the second and third rows, we find that MLOT exhibits mean reversion for non-premium PIPE investors but not for premium investors.  Synthesizing with our prior results, we conclude that premium investors act in a certification role or earn agency rents, whereas non-premium investors serve in an insurance role.

A potential concern with the estimate of $\delta_2$ in column (2) of Table \ref{tab:mlot_last} is that sponsors’ decisions to include non-premium investors may be correlated with unobserved deal quality. Indeed, favoritism implies that the sponsor is more likely to involve non-premium investor(s) for a weak deal. If so, the sample used to identify the parameter is selected and not representative of the population. 
Because the sponsor is likely to allocate higher MLOT to non-premium investors in weak deals, the above estimate can be biased upwards.
 
We use the instrumental variable (IV) approach to address this problem. 
The idea behind our choice of IV is based on the following observation: suppose $v= v_1\times v_2$ is endogenous because $v_1$ is endogenous, then $v_2$ can be an IV for $v$ under additional assumptions. 
For a recent application of this intuition, see \cite{FabraReguant2014} and \cite{AryalCilibertoLeyden2021}. In our setting,  $v$ is  $\text{Non-Premium}\times(\text{lag-MLOT}-\mu)$ and $v_1$ is the $\text{Non-Premium}$ dummy, and $(\text{lag-MLOT}-\mu)$ is  uncorrelated with the quality of the deal. 

However, $(\text{lag-MLOT}-\mu)$ may not satisfy the exclusion restriction because it directly affects the outcome variable. 
Based on the intuition above, we propose to use mean deviation from the second-to-last deal, i.e., $(\text{lag$^2$-MLOT}-\mu)$ as an excluded IV. Lagged variables are widely used as IVs in dynamic panel data settings under the Arellano-Bond estimator. Furthermore, to capture the nonlinearity of the interaction terms, we include $\{(\text{lag-MLOT}-\mu) \times (\text{lag$^2$-MLOT}-\mu), (\text{lag$^2$-MLOT}-\mu)^2, (\text{lag-MLOT}-\mu)^2\}$ as additional excluded variables.

The results from this estimation are in column (3) of Table \ref{tab:mlot_last}. The first-stage Cragg-Donald Wald F-statistic of 92.82 suggests that our excluded variables are not {weak} instruments. Comparing these estimates with the ones in column (2), we see that, as expected, the fixed-effect estimate for the non-premium investor is biased upwards. 
More importantly, and consistent with the quid pro quo hypothesis, we find that one dollar higher mean deviation of MLOT in the last deal leads to a \$2.729 reduction in the mean deviation from the current deal and vice versa presumably because now the non-premium investor is ``asked'' to participate in tepid deals that have smaller underpricing.  
In contrast, for the premium PIPE investors, the effect is the opposite: if the investor made one more dollar in MLOT above the mean in the last deal, on average, the MLOT in the next deal with the same sponsor increases by 23 cents consistent with the hypothesis that premium PIPE investors generate value-relevant information.\footnote{If a deal has more than one sponsor, we treat each sponsor separately. We verify that the results are qualitatively similar if we exclude SPACs with multiple sponsors. } 

\subsubsection{Allocations and Past Relationship}
Next, we test the Hypothesis \ref{hyp:insurance}-3. A sponsor builds a relationship with non-premium investors by providing MLOT from good deals, so if and when the new target firm is weak, it expects the investors with high past MLOT, i.e., the \emph{relationship investors}, to return the favor by investing in the current deal. 
To this end, we evaluate if a sponsor is more likely to use non-premium investors if their MLOT from the sponsor is high and the target is weak.  

Although we do not observe the quality of the target, we propose to proxy it by the difference between the SPAC's offer share price, i.e., the share price at the SPAC IPO, and the share price on the closing day of the merger. The larger the gap, the lower the quality. 

Then, we estimate the probability that a sponsor allocates securities to a non-premium investor as a function of the past MLOT and a measure of cold, among other variables. 
For a sponsor $s$, let ${\mathcal R}_s$ denote the set of {relationship investors} as the set of all non-premium investors that invested in at least one SPAC with sponsor $s$. 
Let $\text{Participation}_{p, s, i}\in\{0,1\}$ be a binary variable equal to one if investor $p\in {\mathcal R}_{s}$ is allocated securities from SPAC $i$ started by sponsor $s$, and zero otherwise. Then we estimate the participation probability, $\Pr(\text{Participation}_{p, i, s}=1\mid \mathbf{Z}_{p, s, i}^{\top} \boldsymbol{\delta})$, under the assumption that it takes a logit form, with 
\begin{eqnarray}
\mathbf{Z}_{p, s, i}^{\top} \boldsymbol{\delta}&=&\mathbf{X}^\top\boldsymbol{\delta_0}+\delta_1\times \log(1+\text{MLOT(past)}_{p, s})+\delta_2\times\text{Cold}_{i}\notag\\&&+\delta_3\times \log(1+\text{MLOT(past)}_{p,s})\times\text{Cold}_{i},\label{eq:participation}
\end{eqnarray}
where ${\bf X}$ is the vector of controls shown in Table \ref{tab:mlot_last}; 
$\log(1+\text{MLOT(past)}_{p,s})$ is the log of the total MLOT $p$ made in past deals with $s$ and measures the past relationship between $s$ and $p$; and, and as we defined above, $\text{Cold}_i$ is the measure of the quality of the deal $i$. 

We estimate (\ref{eq:participation}) using the method of maximum likelihood and present the results in column (1) of Table \ref{tab:relationship}. The estimated coefficient of the interaction term (third row) in column (1) is positive (p-value $<0.01$), supporting the quid pro quo hypothesis that the sponsor first builds a relationship and draws it down as and when needed.

\begin{table}[t!]
\centering
\caption{Sponsor and Non-Premium PIPE Investors}
\centering
\scalebox{0.9}{
\begin{tabular}{l*{5}{c}}
\toprule
                    &\multicolumn{4}{c}{Participation}&\multicolumn{1}{c}{Allocation}\\
                    &\multicolumn{1}{c}{(1)}&\multicolumn{1}{c}{(2)}&\multicolumn{1}{c}{(3)}&\multicolumn{1}{c}{(4)}&\multicolumn{1}{c}{(5)}\\
\midrule
$\log(\text{1+MLOT(past)})$&       0.085         &      -0.004         &      -0.088         &      -0.335$^{***}$&       0.019 \\
                    &     (0.136)         &     (0.123)         &     (0.098)         &     (0.081)         &     (0.010)         \\
                    &&&&&[0.013]\\
Cold&       0.006$^{***}$&       0.004$^{***}$&      -0.003         &       0.003$^{**}$ &      -0.00003         \\
                    &     (0.001)         &     (0.001)         &     (0.002)         &     (0.002)         &     (.00006)         \\
                    &&&&&[0.0006]\\
$\log(\text{1+MLOT(past)}) \times \text{Cold}$&       0.008$^{***}$&       0.011$^{***}$&       0.011$^{***}$&       0.008$^{***}$&       0.0003$^{***}$\\
                    &     (0.002)         &     (0.003)         &     (0.003)         &     (0.002)         &     (0.0001)        \\
                    &&&&&[0.00012]\\

Outside Option&                     &      -0.017$^{***}$&      -0.007$^{***}$&      -0.002         &                     \\
                    &                     &     (0.003)         &     (0.002)         &     (0.002)         &                     \\
$\widehat{\mathbb{E}(\text{Participation}\mid {\mathbf Z})}$   &                     &                     &                     &                     &       0.239$^{***}$\\
                    &                     &                     &                     &                     &     (0.018)         \\
                    &&&&&[0.024]\\
\midrule
Log IPO proceeds            & \xmark & \checkmark & \checkmark & \checkmark & \checkmark \\
Days Remaining              & \xmark & \checkmark & \checkmark & \checkmark & \checkmark \\
No. Premium Investors       & \xmark & \checkmark & \checkmark & \checkmark & \xmark \\
Sponsor Dummies             & \xmark & \xmark     & \checkmark & \checkmark & \xmark \\
Month Dummies               & \xmark & \xmark     & \xmark     & \checkmark & \checkmark \\
\midrule
Pseudo/Adj. $ R^2 $        &   0.0098                  &       0.0644              &      0.2763               &       0.4061              &      0.8114               \\
Obs.                &      10,104         &      10,044         &       7,917         &       7,913         &       7,838         \\
\bottomrule
\end{tabular}}%
\label{tab:relationship}
\begin{figurenotes}
Columns (1) to (4) of this table show the MLE of participation (\ref{eq:participation}), and column (5) shows the estimates of the allocation (\ref{eq:share}). Each observation is at (SPAC, sponsor, non-premium PIPE investor) level. For each SPAC, non-premium investors include all relationship investors who participate in at least one SPAC deal initialized by the sponsor during the sample period. For the first four columns, the dependent variable, $Participation$, equals one if the non-premium investor participates in the focal SPAC deal.  
The dependent variable for the last column, $Allocation$, is the investor's allocation as a fraction of all relationship investors' allocations. 
The key independent variables include the non-premium investor's MLOT earned with this sponsor in the past in log form, $\log(1+ \text{MLOT ({past})})$, and a measure of the (reciprocal) of the deal quality, Cold. For the allocation regression, we also include the estimated probability of participation using estimates from column (4) as an additional variable. 
All continuous variables at winsorized at $99^{th}$ percentile.
Robust standard errors are clustered at the sponsor-PIPE pair level and reported in parentheses. For column (5), Bootstrap standard errors (500 replications) are reported in the square brackets. $^*, ^{**}$, and $^{***}$ denote \textit{p}-values less than 0.1, 0.05, and 0.01, respectively.
\end{figurenotes}
\end{table}

One possible concern with the estimation is that it ignores the sponsor's outside option, which may be correlated with the unobserved error. A better outside option lowers the probability of issuing securities to a non-premium investor and vice versa. 
Ignoring outside options could introduce attenuation bias if cold deals have profitable outside options. 
To alleviate this concern, we use the highest MLOT among all premium PIPE investors \emph{not included} in the current deal as an additional control variable to capture the strength of the outside option. We also include the log of IPO proceeds, days remaining, and the number of premium investors as additional covariates. 

The estimates from this new model are in column (2) of Table \ref{tab:relationship}. 
A better outside option lowers the probability of selecting the non-premium investor, and our previous finding becomes slightly stronger. The results are robust when we include dummy variables for the sponsor and the month when the deal is announced. In all cases, the coefficient for the interaction term is positive and estimated precisely (p-value $<0.01$).

To help interpret these estimates, we evaluate the probability in (\ref{eq:participation}) using the estimates from column (4) of Table \ref{tab:relationship} at different levels of past MLOT and quality of the target, and hold other variables at their average value. 
Investors with a stronger relationship have higher participation rates in deals with lower demand. 
In SPAC deals with Cold at its $90^{th}$-percentile, non-premium investors who have a strong relationship with the sponsor (one standard deviation above the mean) are 3.54 pps more likely to participate than in SPAC deals with Cold at its $10^{th}$-percentile. Conversely, non-premium investors without a prior relationship with the sponsor are less likely to participate in cold deals.

Next, we examine if, conditional on participation, the sponsor allocates a higher portion of securities to non-premium investors with a stronger relationship when a deal is cold and vice versa. To this end, we determine the allocation of SPAC $i$'s securities from sponsor $s$ to PIPE investor $p$ among all relationship investors ${\mathcal R}_s$, $\text{Allocation}_{p, s, i}\in[0,1]$, as the ratio of $p$'s shares to the total shares allocated to ${\mathcal R}_s$. 
We set the share to zero if a non-premium investor does not participate in a deal.  
The mean allocation is $2.93$, and the standard deviation is $8.77$. 
Furthermore, the allocation is skewed towards zero, with the median of $0$ and the $99^{th}$-percentile at $42.37$. 
There are 1,792 non-premium investors, and the allocation has a mass at zero, so we estimate the following Tobit model:
\begin{eqnarray}
\begin{cases}\text{Allocation}_{p, s, i}^*=&\mathbf{W}_{p, i, s}^\top \boldsymbol{\gamma^*} + u_s,\\
\text{Allocation}_{p, s, i}=&\min\{\max\{0,\mathbf{W}_{p, i, s}^\top \boldsymbol{\gamma} + \nu_1\},1\}\\
\mathbf{W}_{p, i, s}^{\top} \boldsymbol{\gamma}=&\mathbf{X}^\top\boldsymbol{\gamma_0}+\gamma_1\times \log(1+\text{MLOT(past)}_{p,s})+\gamma_2\times\text{Cold}_{i}\\&+\gamma_3\times \log(\text{MLOT(past)}_{p,s}+1)\times\text{Cold}_{i}+\gamma_4\times\widehat{\mathbb{E}(\text{Participation}|{\mathbf Z})}_{p, s, i},
\end{cases}\label{eq:share}
\end{eqnarray}
where $(\mathbf{W}, \text{Allocation}_{p, s, i})$ is always observed but the latent variable $\text{Allocation}_{p, s, i}^*$ is observed only when $\text{Allocation}_{p, i, s} \in[0,1]$. We further assume that $\mathbf{W}$ is independent of $(u_1, ]\nu_1)$ and that $\nu_1$ is a mean-zero Normal random variable with unknown variabce and $\mathbb{E}(u_1\mid \nu_1)=\rho \nu_1$.
Furthermore, $\widehat{\mathbb{E}(\text{Participation}\mid {\mathbf Z})}_{p, s, i}$ is the estimated participation probability of allocation given ${\bf Z}$ and is determined by the estimates in column (4) of Table \ref{tab:relationship}. Thus, the outside option serves as our excluded variable, with the underlying exclusion restriction being that the outside option should not directly affect a non-premium investor's allocation.

In Column (5) of Table \ref{tab:relationship}, the estimates indicate a positive association between a higher probability of participation and an increased allocation of shares. 
Moreover, non-premium investors who have previously profited from deals with the sponsor receive a larger share of allocation than their counterparts when a deal is considered cold. 

In economic terms, a one standard deviation increase in the interaction term leads to a 5.8\% increase in allocation relative to its standard deviation, amounting to 14.80\% of the sample mean. In our sample, an average non-premium investor is allocated 2.68\% of the total non-premium investors' allocations. Furthermore, as the non-premium investor's MLOT (past) times the deal's Cold measure increase together by one standard deviation, the allocation fraction rises by 0.40 pps to 3.08\%. In terms of dollar amount, the allocation increases from \$0.40 million to \$0.46 million.
These findings support the hypothesis that non-premium investors and sponsors engage in quid pro quo agreements.

\subsubsection{Relationship Capital and Liquidation Risk}
Our estimates so far suggest that the relationship investors, who are also non-premium investors, support sponsors by participating in weak deals. 
These relationship investors received larger MLOTs from the same sponsor in the previous deals than other non-premium investors.  
It stands to reason that a sponsor builds these relationships with an eye toward weak deals in the future--as insurance. 
If so, a sponsor with more ``relationship capital'' can afford to take more risk by working with weak targets.

Next, we test the Hypothesis \ref{hyp:insurance}-4 by estimating the liquidation probability as a function of MLOT earned by non-premium investors in \emph{previous} deals initiated by the same sponsor using a linear probability model. 

In our sample, there are 191 liquidations and 473 successful mergers. We verify if the sponsor's ``relationship payments'' to non-premium investors in the SPAC lowers the probability of liquidation.
To this end, we use a linear probability model in which the dependent variable is a binary variable equal to one if the SPAC is liquidated and zero otherwise. The explanatory variables include past MLOTs the non-premium investors have made from the current sponsor and several other variables. 

The results from this estimation exercise are shown in Table \ref{tab:relationship_capital_liquidation}. Column (1) indicates SPACs initiated by sponsors with stronger relationships with non-premium investors are less likely to liquidate.
We progressively introduce additional control variables from columns (2) to (6) as shown in Table \ref{tab:relationship_capital_liquidation}. Column (2) includes the SPAC's IPO month fixed effect to account for time differences. Column (3) incorporates fixed effects for the SPAC's targeted sector at its IPO to capture systematic sectoral differences that may affect liquidation risk. Column (4) includes additional jurisdiction fixed effects based on the SPAC's registration to capture legal system differences that may influence liquidation rules. Column (5) includes the sponsor fixed effect estimated from Equation (\ref{eq:akm}) as a proxy for sponsor quality. The estimates show that SPAC's liquidation risk decreases if its sponsor has a higher relationship with non-premium investors, and this effect is stable across different specifications. 

\begin{table}[!t]
\centering
\caption{Liquidation Risk and Relationship}
\centering
\scalebox{0.88}{
\hspace{-2mm}
\begin{tabular}{l*{6}{c}}
\toprule
                    &\multicolumn{1}{c}{(1)}&\multicolumn{1}{c}{(2)}&\multicolumn{1}{c}{(3)}&\multicolumn{1}{c}{(4)}&\multicolumn{1}{c}{(5)}&\multicolumn{1}{c}{(6)}\\

\midrule
Past Relationship (Non-premium)  &      -0.064\sym{**} &      -0.074\sym{**} &      -0.077\sym{**} &      -0.079\sym{***}&      -0.081\sym{***}&      -0.065\sym{**} \\
                        &     (0.029)         &     (0.030)         &     (0.030)         &     (0.030)         &     (0.031)         &     (0.030)         \\
Past Relationship (Premium)      &       0.057\sym{*}  &       0.049         &       0.054\sym{*}  &       0.059\sym{*}  &       0.060\sym{*}  &       0.054\sym{*}  \\
                        &     (0.029)         &     (0.033)         &     (0.032)         &     (0.033)         &     (0.033)         &     (0.033)         \\
SPAC Return           &                     &                     &                     &                     &                     &      -0.0001\sym{*}  \\
                        &                     &                     &                     &                     &                     &     (0.00005)         \\
$\mathbbm{1}$\{Premium Participation\}            &                     &                     &                     &                     &                     &      -0.372\sym{***}\\
                        &                     &                     &                     &                     &                     &     (0.036)         \\
\midrule
Controls
& \checkmark                 & \checkmark                 & \checkmark                 & \checkmark                 & \checkmark & \checkmark                 \\
IPO Month FEs
& \xmark                  & \checkmark                 & \checkmark                 & \checkmark                 & \checkmark    & \checkmark             \\
Target Sector FEs
& \xmark                  & \xmark                  & \checkmark                 & \checkmark                 & \checkmark      & \checkmark           \\
Jurisdiction FEs
& \xmark                 & \xmark                  & \xmark                  & \checkmark                 & \checkmark       & \checkmark          \\
Sponsor Quality
& \xmark                  & \xmark                  &\xmark                 & \xmark                  & \checkmark         & \checkmark        \\
\midrule
Adj. $ R^2 $        &       0.048         &       0.121         &       0.126         &       0.126         &       0.125         &       0.254         \\
Obs.                &         806         &         781         &         758         &         757         &         757         &         757         \\
\bottomrule
\end{tabular}}%
\label{tab:relationship_capital_liquidation}
\begin{figurenotes}
This table shows that SPACs initiated by sponsors with stronger relationships with non-premium investors are less likely to liquidate. All regressions use linear probability models. The dependent variable is a dummy that equals one if the SPAC liquidates and zero otherwise. The main independent variables are the total MLOT earned by premium and non-premium investors in previous SPACs initiated by the same sponsor.
SPAC Return is the return of holding the SPAC's share from the offering date to one day before the merger completion date or the liquidation date. $\mathbbm{1}$\{Premium Participation\} is a dummy variable that equals one if any premium investors participate in the focal deal.
Controls include the (log) IPO proceeds, the number of past successful deals, a dummy variable for the presence of PIPE (pre) investors, and the number of active SPACs. Sponsor quality is measured by the sponsor's fixed effect estimated from (\ref{eq:akm}).
All continuous variables are winsorized at the $1^{st}$ and $99^{th}$ percentiles. Robust standard errors, clustered at the sponsor level, are reported in parentheses. $^*, ^{**}$, and $^{***}$ denote \textit{p}-values less than 0.1, 0.05, and 0.01, respectively.
\end{figurenotes}
\end{table}

A possible threat to the identification is that the set of non-premium investors could be endogenous if it is correlated with the unobserved quality of the target. If so, the estimated negative effect could be due to the positive selection between target and sponsor in that high-quality targets work with high-quality sponsors, who likely have completed more SPAC deals in the past and, therefore, have higher relationships with non-premium investors. So, the liquidation risk is lower because of the quality of the target. We include additional controls correlated with the target's quality to address this issue. In particular, column (6) controls for the market return before the SPAC's completion or liquidation and includes a dummy variable equal to one if a premium investor participates in the deal. Even with these new controls, we find that the estimated effect is negative and statistically significant at 5\% confidence level.\footnote{
Although not reported, we verify that our finding is robust to using the competing-risk model to account for right censoring because 16\% of the announced deals in our sample are still searching for a target.} In particular, the estimate suggests that doubling the sponsor's relationship payments to non-premium investors decreases liquidation probability by 6.5\%, which is 28.2\% of the sample average liquidation rate, and this estimated effect is statistically significant at the 5\% confidence level.\footnote{On March 30, 2022, the SEC proposed new rules for SPAC mergers that made mergers more onerous and, consequently, relationships with non-premium investors more important. 
Indeed, estimating the specification in Table \ref{tab:relationship_capital_liquidation}, column (6), separately for those SPACs that announced target before the law change and for those after, we find that the coefficient increases from $-0.04$ to $-0.11$ but are statistically indistinguishable.}

In summary, we find evidence consistent with these conditions: (i) premium PIPE investors generate value-relevant information, (ii) non-premium PIPE investors are paid agency fees, and (iii) the non-premium investors help close weaker deals in the future.
In Appendix \ref{section:robustness}, we assess the robustness of our results with respect to removing outlier investors, placebo treatments, and the definition of premium PIPE investors.

\section{Conclusion}

In this paper, we consider the relationship between intermediaries and investors in the securities issuance market. Our analysis distinguishes between premium and non-premium investors, revealing distinct roles for each investor type. Premium investors act in a certification or agency manner, whereas non-premium investors engage in relationships with intermediaries as \cite{LJUNGQVIST2007375} hypothesized. Specifically, when non-premium investors are allocated shares of hot deals, they are expected to reciprocate by supporting future lukewarm deals. These relationships serve as insurance for both the issuer and the intermediary.

We examine this insurance hypothesis in the context of SPACs, where the sponsor serves as an intermediary between the selling firm and PIPE investors. We focus on this setting because of its data advantages over traditional IPOs. With SPACs, we can directly observe stock price responses to public announcements, including the identity of the target and the participation decisions of PIPE investors.  Additionally, we can directly observe the returns earned by each PIPE in each deal, allowing us to form a comprehensive panel data set of sponsor-PIPE relationships. Such sponsor-PIPE relationships data are unavailable in traditional IPOs, because there the initial allocations are proprietary information and investors often sell shares before filing a 13F, if at all.

Our findings indicate that premium investors are associated with higher announcement period returns, less redemption, and a higher likelihood of successful SPAC mergers. Conversely, non-premium investors exhibit mean-reverting returns, indicative of relationship capital buildup and drawdown.  To establish the channel through which relationships are drawn down, we examine the participation and allocation of PIPE investors during cold deals, and show that these measures rise with past high returns.
Independently of deal quality, sponsor-investor relationships significantly reduce the liquidation probability of current deals, demonstrating an insurance benefit accruing to intermediaries and issuers.

\bibliographystyle{aea}
\bibliography{literature.bib}

@article{abowd1999high,
  title={High Wage Workers and High Wage Firms},
  author={Abowd, John M and Kramarz, Francis and Margolis, David N},
  journal={Econometrica},
  volume={67},
  number={2},
  pages={251--333},
  year={1999}}

@article{ROCK1986,
title = {Why new issues are underpriced},
journal = {Journal of Financial Economics},
volume = {15},
number = {1},
pages = {187-212},
year = {1986},
issn = {0304-405X},
doi = {https://doi.org/10.1016/0304-405X(86)90054-1},
author = {Kevin Rock},
}

@article{KlausnerOhlroggeRuan2022,
	author = {Michael Klausner and Michael Ohlrogge and Emily Ruan},
	journal = {Yale Journal of Regulation},
	number = {1},
	pages = {228-303},
	title = {A Sober Look at SPACs},
	volume = {39},
	year = {2022}}

@article{JenkJones,
	author = {Jenkinson, Tim and Jones, Howard and Suntheim, Felix},
	doi = {https://doi.org/10.1111/jofi.12703},
	eprint = {https://onlinelibrary.wiley.com/doi/pdf/10.1111/jofi.12703},
	journal = {Journal of Finance},
	number = {5},
	pages = {2303-2341},
	title = {Quid Pro Quo? What Factors Influence IPO Allocations to Investors?},
	url = {https://onlinelibrary.wiley.com/doi/abs/10.1111/jofi.12703},
	volume = {73},
	year = {2018}}

@article{ChemmanueHuHuang2010,
	author = {Thomas J. Chemmanur and Gang Hu and Jiekun Huang},
	journal = {Review of Financial Studies},
	number = {12},
	pages = {4496-4540},
	title = {The Role of Institutional Investors in Initial Public Offerings},
	volume = {23},
	year = {2010}}

@article{RitterWelch2002,
	author = {Jay R. Ritter and Ivo Welch},
	journal = {Journal of Finance},
	number = {4},
	pages = {1795-1828},
	title = {A Review of IPO Activity, Pricing, and Allocations},
	volume = {57},
	year = {2002}}

@article{gofman2022spacs,
	author = {Gofman, Michael and Yao, Yuchi},
	journal = {Available at SSRN 4148668},
	title = {SPACs' Directors Network: Conflicts of Interest, Compensation, and Competition},
	year = {2022}}

@article{AltiCohn,
	author = {Alti, Aydogan and Jonathan Cohn},
	title = {A Model of Informed Intermediation in the Market for Going Public},
	year = {2023}}

@article{Ince,
	author = {{\.I}nce, \"{O}zg\"{u}r \c{S}.},
	journal = {The Quarterly Journal of Finance},
	number = {03},
	pages = {1450009},
	title = {Why Do IPO Offer Prices Only Partially Adjust?},
	volume = {04},
	year = {2014}}

@article{FUDENBERG,
	author = {Drew Fudenberg and David K. Levine},
	journal = {Journal of Economic Theory},
	number = {1},
	pages = {103-135},
	title = {Efficiency and Observability with Long-Run and Short-Run Players},
	volume = {62},
	year = {1994}}

@article{FabraReguant2014,
	author = {Natalia Fabra and Mar Reguant},
	journal = {American Economic Review},
	number = {9},
	pages = {2872-2899},
	title = {Pass-Through of Emissions Costs in Electricity Markets},
	volume = {104},
	year = {2014}}

@article{AryalCilibertoLeyden2021,
	author = {Gaurab Aryal and Federico Ciliberto and Benjamin T. Leyden},
	journal = {Review of Economic Studies},
    number = {6},
	pages = {3055-3084},
	title = {Coordinated Capacity Reductions and Public Communication in the Airline Industry},
    volume ={89},
	year = {2022}}

@article{BenvenisteLjungqvistWilhelmYu2003,
	author = {Lawrence M. Benveniste and Alexander Ljungqvist and Wilhelm Jr, William J. and Xiaoyun Yu},
	journal = {Journal of Finance},
	number = {2},
	pages = {577-608},
	title = {Evidence of Information Spillovers in the Production of Investment Banking Services},
	volume = {58},
	year = {2003}}

@article{BradleyJordan2002,
	author = {Daniel J. Bradley and Bradford Jordan},
	journal = {Journal of Financial and Quantitative Analysis},
	number = {4},
	pages = {595-616},
	title = {Partial Adjustment to Public Information and IPO Underpricing},
	volume = {37},
	year = {2002}}

@article{WangYung2011,
	author = {Wei Wang and Chris Yung},
	journal = {Review of Finance},
	number = {2},
	pages = {301-325},
	title = {IPO Information Aggregation and Underwriter Quality},
	volume = {15},
	year = {2011}}

@article{LoughranRitter2002,
	author = {Tim Loughran and Jay R. Ritter},
	journal = {Review of Financial Studies},
	number = {2},
	pages = {413-444},
	title = {Why Don't Issuers Get Upset about Leaving Money on the Table in IPOs?},
	volume = {15},
	year = {2002}}

@article{LoughranRitter2004,
	author = {Tim Loughran and Jay R. Ritter},
	journal = {Financial Management},
	number = {3},
	pages = {5-37},
	title = {Why Has IPO Underpricing Changed over Time?},
	volume = {33},
	year = {2004}}

@article{BeattyRitter1986,
	author = {Randolph P. Beatty and Jay R. Ritter},
	journal = {Journal of Financial Economics},
	number = {1-2},
	pages = {213-232},
	title = {Investment Banking, Reputation, and the Underpricing of Initial Public Offerings},
	volume = {15},
	year = {1986}}

@book{CampbellLoMacKinlay1997,
	author = {John Y. Campbell and Andrew W. Lo and A. Craig MacKinlay},
	publisher = {Princeton University Press},
	title = {The Econometrics of Financial Markets},
	year = {1997}}

@article{benveniste1989investment,
	author = {Benveniste, Lawrence M and Spindt, Paul A},
	journal = {Journal of Financial Economics},
	number = {2},
	pages = {343--361},
	publisher = {Elsevier},
	title = {How Investment Bankers Determine the Offer Price and Allocation of New Issues},
	volume = {24},
	year = {1989}}

@article{hanley1993underpricing,
	author = {Hanley, Kathleen Weiss},
	journal = {Journal of Financial Economics},
	number = {2},
	pages = {231--250},
	publisher = {Elsevier},
	title = {The Underpricing of Initial Public Offerings and the Partial Adjustment Phenomenon},
	volume = {34},
	year = {1993}}

@article{lowry2017initial,
  title={Initial public offerings: A synthesis of the literature and directions for future research},
  author={Lowry, Michelle and Michaely, Roni and Volkova, Ekaterina and others},
  journal={Foundations and Trends{\textregistered} in Finance},
  volume={11},
  number={3-4},
  pages={154--320},
  year={2017},
  publisher={Now Publishers, Inc.}
}

@article{booth1986capital,
	author = {Booth, James R and Smith II, Richard L},
	journal = {Journal of Financial Economics},
	number = {1-2},
	pages = {261--281},
	publisher = {Elsevier},
	title = {Capital Raising, Underwriting and the Certification Hypothesis},
	volume = {15},
	year = {1986}}

@article{smith1986investment,
	author = {Smith Jr, Clifford W},
	journal = {Journal of Financial Economics},
	number = {1-2},
	pages = {3--29},
	publisher = {Elsevier},
	title = {Investment Banking and the Capital Acquisition Process},
	volume = {15},
	year = {1986}}

@article{ritter2021spacs,
	author = {Gahng, Minmo and Ritter, Jay R. and Zhang, Donghang},
	journal = {Review of Financial Studies},
	title = {SPACs},
	year = {2024}}

@article{LJUNGQVIST2007375,
title = {IPO Underpricing},
editor = {B. Espen Eckbo},
booktitle = {Handbook of Empirical Corporate Finance},
publisher = {Elsevier},
address = {San Diego},
pages = {375-422},
year = {2007},
series = {Handbooks in Finance},
issn = {15684997},
doi = {https://doi.org/10.1016/B978-0-444-53265-7.50021-4},
url = {https://www.sciencedirect.com/science/article/pii/B9780444532657500214},
author = {Alexander Ljungqvist}
}

@ARTICLE{BenvenisteBusaba,
title = {Information Externalities and the Role of Underwriters in Primary Equity Markets},
author = {Benveniste, Lawrence M. and Busaba, Walid Y. and Wilhelm, William },
year = {2002},
journal = {Journal of Financial Intermediation},
volume = {11},
number = {1},
pages = {61-86},
url = {https://EconPapers.repec.org/RePEc:eee:jfinin:v:11:y:2002:i:1:p:61-86}
}
\newpage
\clearpage

\setcounter{section}{0}
\setcounter{equation}{0}
\setcounter{table}{0}
\renewcommand{\thesection}{A.\arabic{section}}
\renewcommand{\thesubsection}{A.\arabic{subsection}}
\renewcommand{\theequation}{A.\arabic{equation}}
\renewcommand{\thetable}{A\arabic{table}}
\renewcommand{\thefigure}{A\arabic{figure}}

\appendix
\begin{center}
{\Large\bf APPENDIX}
\end{center}

\section{Robustness\label{section:robustness}} 
In Section \ref{section:results}, we found evidence consistent with the hypotheses that premium investors produce value-relevant information and non-premium investors engage in quid pro quo arrangements. We also found evidence that the quid pro quo relationship between a sponsor and non-premium investors enables weaker private firms to go public.
In this section, we perform a series of robustness exercises.

\subsection{\bf Possible Outliers}

\begin{figure}[ht!]
\caption{Scatter Plots of SPACs}
\hspace{-0.4in}\includegraphics[scale=0.3]{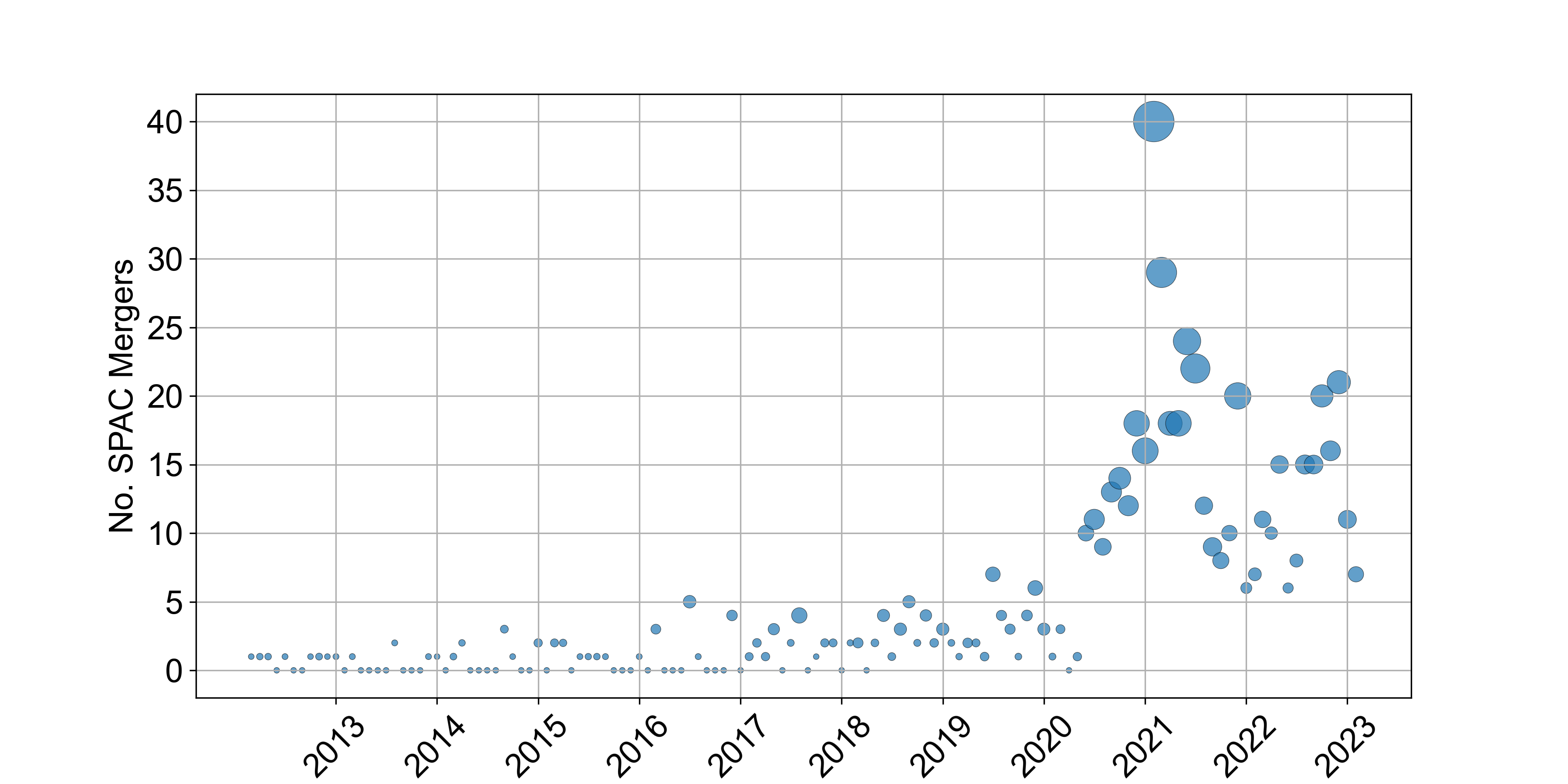}~
\includegraphics[scale=0.3]{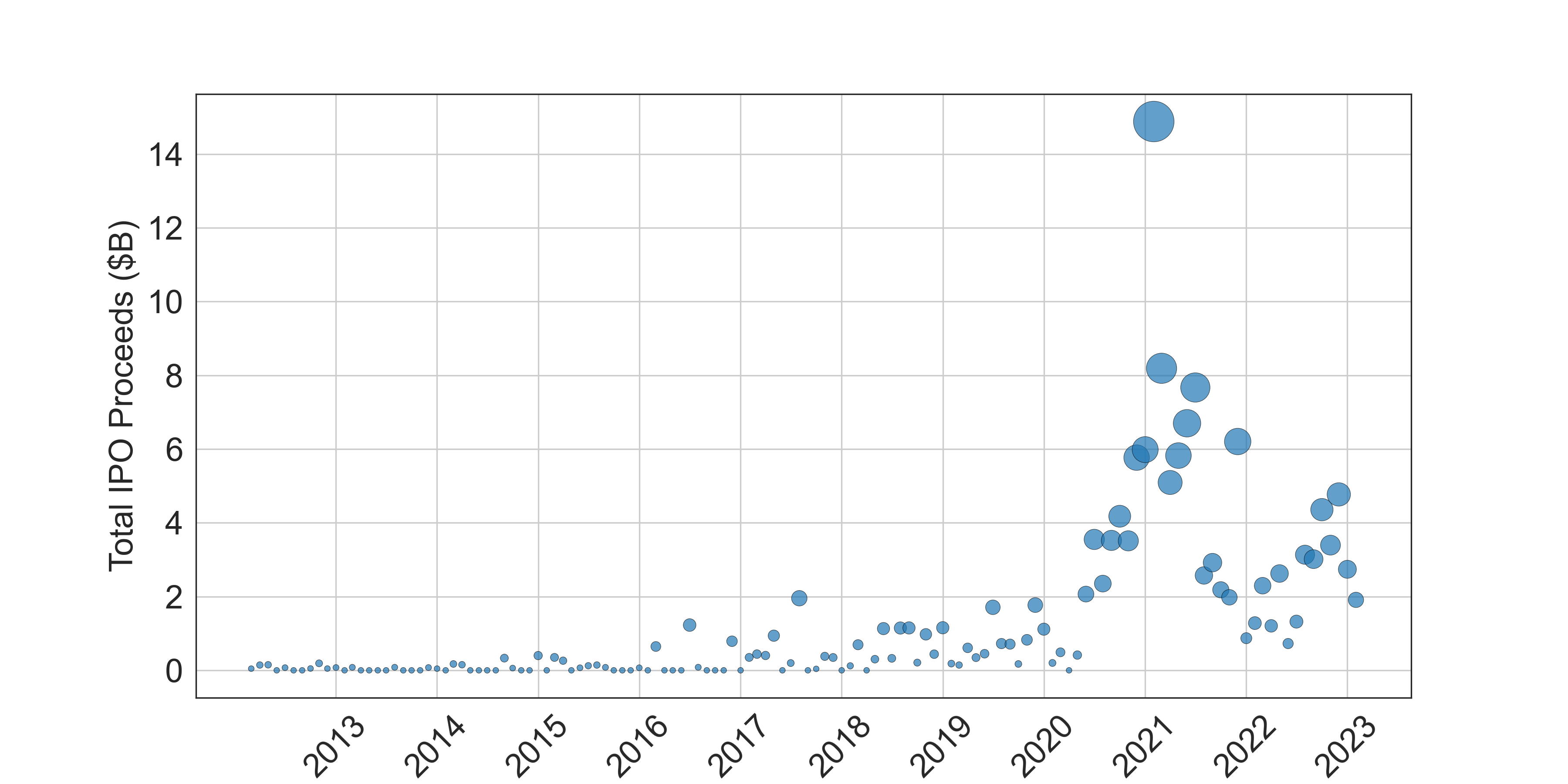}
\label{fig:size_number_spacs}
\begin{figurenotes}
The figure shows the scatter plot of SPACs by month and year. The top figure shows the number of SPAC mergers, and the bottom figure shows the deal sizes (total IPO proceeds) in billions of dollars. The size of each dot is proportional to the number of deals in a month (top) and the total IPO proceeds (bottom) in that month.   
\end{figurenotes}
\end{figure}

As shown in Figure \ref{fig:size_number_spacs}, although our sample starts in 2010, most of the variations come from SPAC from 2020 to 2022. Although our sample includes the so-called boom and bust cycle for SPAC, we may worry about the effects of outliers in our analysis. One such concern is that not all 34 premium PIPE investors (in Table \ref{tab:premium_investor}) produce value-relevant information and that our results are driven by either the large premium PIPE investors or that the small premium PIPE investors are, in fact, non-premium investors. 

To address this concern, we exclude six (large) premium investors with at least \$1 billion invested in SPACs and re-estimate all our models. Then, we repeat the same exercise excluding seven (small) premium investors with at most \$400 million invested in SPACs.

\begin{table}[!t]
\caption{Robustness: Excluding Outliers}
\hspace{-0.83in}\scalebox{0.9}{
\begin{tabular}{l*{8}{c}}
\toprule
                    &\multicolumn{2}{c}{Redemption}&\multicolumn{2}{c}{Announcement-day}&\multicolumn{2}{c}{Mean Reversion}&\multicolumn{2}{c}{Liquidation}\\
                    & \multicolumn{2}{c}{Rate (\%)}&\multicolumn{2}{c}{Return (\%)}&\multicolumn{2}{c}{MLOT}\\
                    &(1)&(2)&(3)&(4)&(5)&(6)&(7)&(8)\\
\midrule   
PIPE (pre)             &0.174\sym{*}            &0.151                 &-0.082\sym{*}            &-0.070                    &                &                &                     &                     \\
                       &(0.098)                 &(0.093)               &(0.044)                  &(0.043)                   &                &                &                     &                     \\
Premium PIPE (at)      &-0.244\sym{***}         &-0.219\sym{***}       &0.144\sym{**}            &0.126\sym{***}            &0.153           &0.149           &0.054\sym{*}         &0.050                \\
                       &(0.082)                 &(0.052)               &(0.059)                  &(0.041)                   &(0.104)         &(0.099)         &(0.033)              &(0.033)              \\ 
Non-premium PIPE (at)  &-0.041                  &-0.038                &0.001                    &-0.001                    &-1.633\sym{***} &-2.075\sym{***} &-0.062\sym{**}       &-0.060\sym{**}               \\
                       &(0.031)                 &(0.031)               &(0.014)                  &(0.015)                   &(0.531)         &(0.615)         &(0.029)              &(0.030)              \\
PIPE (post)            &0.152\sym{***}          &0.148\sym{***}        &                         &                          &                &                &                     &                     \\
                       &(0.036)                 &(0.035)               &                         &                          &                &                &                     &                     \\
Exclude Large Premium Investors  &\checkmark              &\xmark                &\checkmark               &\xmark                    &\checkmark      &\xmark          &\checkmark           &\xmark               \\
Exclude Small Premium Investors  &\xmark                  &\checkmark            &\xmark                   &\checkmark                &\xmark          &\checkmark      &\xmark               &\checkmark           \\
\midrule
Pseudo/Adj. $ R^2 $    &0.386                   &0.397                 &0.069                    &0.090                     &                &                &0.254                &0.253               \\
F-stat (first-stage)   &                        &                      &                         &                          &175.54          &140.73          &                     &                     \\
Observations           &430                     &430                   &417                      &417                       &3,487           &3,578           &758                  &757                  \\
\bottomrule
\end{tabular}}%
\label{tab:t_iv2_r_outlier}
\begin{figurenotes}
This table presents estimates after excluding outliers in premium investors. 
Premium PIPE (at) investors are listed in Table \ref{tab:premium_investor}.
They include Fidelity Management \& Research Company, BlackRock, Inc., Capital Research and Management Company, Alyeska Investment Group L.P., Morgan Stanley, and Millennium Management, LLC.
Small premium investors are those that invested less than \$0.4 billion and participated in at most 50 SPAC deals. 
They include Ghisallo Capital Management LLC, Jane Street Group, LLC, Park West Asset Management LLC, Kepos Capital LP, Linden Advisors LP, Schonfeld Strategic Advisors LLC, and BlueCrest Capital Management Ltd.
Columns (1) and (2) have the same specification as Table \ref{tab:t_redemption_premium}, column (5).
Columns (3) and (4) have the same specification as Table \ref{tab:t_r_ann_premium}, column (5).
Columns (5) and (6) have the same specification as Table \ref{tab:mlot_last}, column (3).
Columns (7) and (8) have the same specification as Table \ref{tab:relationship_capital_liquidation}, column (6).
$^*, ^{**}$, and $^{***}$ denote \textit{p}-values less than 0.1, 0.05, and 0.01, respectively. 
\end{figurenotes}
\end{table}

The results are presented in Table \ref{tab:t_iv2_r_outlier}. In particular, the estimates from excluding large premium investors are in columns (1), (3), (5), and (7) of Table \ref{tab:t_iv2_r_outlier}, and the estimates from excluding small premium investors are in the remaining columns. 
Comparing these estimates with those in columns (5), (5), (3), and (6) of Tables \ref{tab:t_redemption_premium}, \ref{tab:t_r_ann_premium}, \ref{tab:mlot_last}, and \ref{tab:relationship_capital_liquidation}, respectively, we see that our main findings are neither due only to the large or small premium investors.

\subsection{\bf Placebo Exercises}

\begin{figure}[h]
\centering
\caption{``Placebo'' Exercises }
\begin{subfigure}[b]{0.5\textwidth}
\includegraphics[scale=0.3]{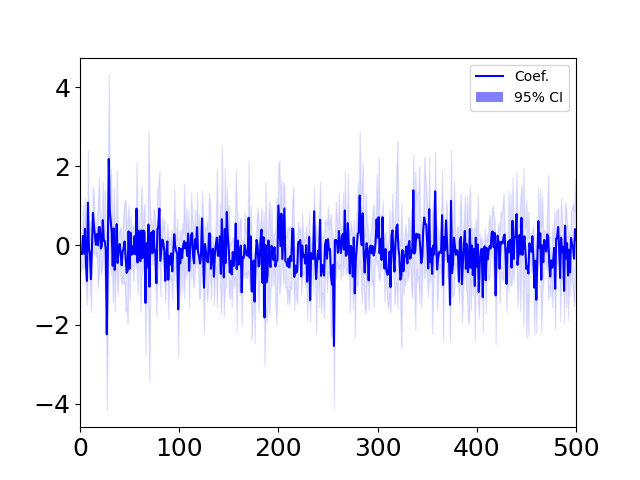}~
\includegraphics[scale=0.3]{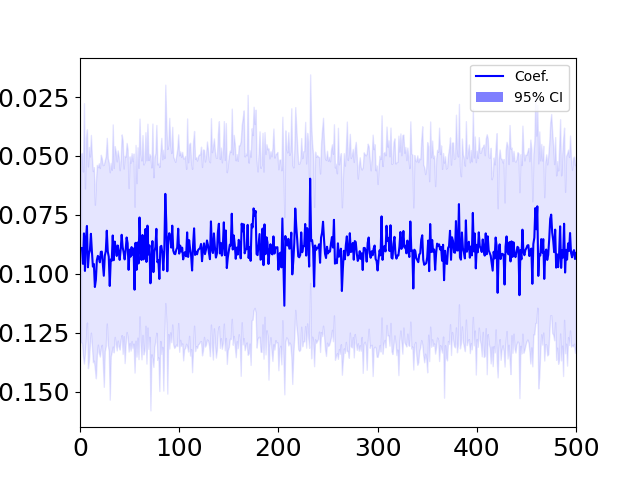}
\end{subfigure}
\caption*{(i) Redemption Rates: Premium (left) and Non-Premium(right).}

\begin{subfigure}[b]{0.5\textwidth}
\includegraphics[scale=0.3]{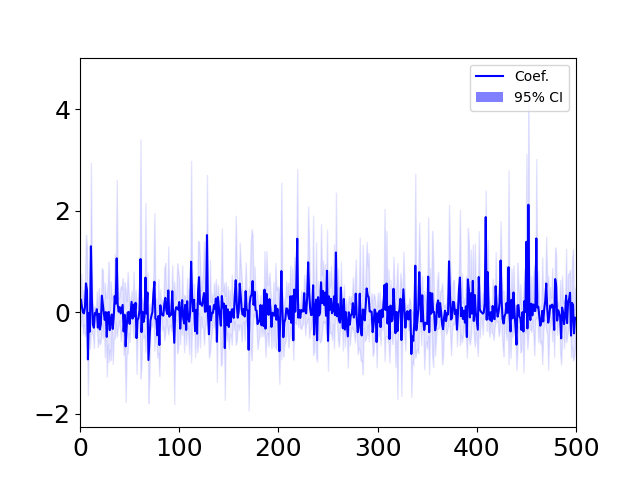}~
\includegraphics[scale=0.3]{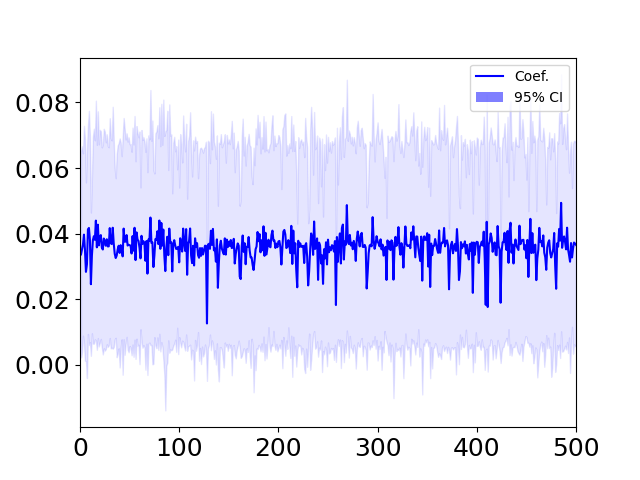}
\end{subfigure}
\caption*{(ii) Announcement-day Returns: Premium (left) and Non-Premium (right)}
\begin{subfigure}[b]{0.5\textwidth}
\vspace{-0.25cm} 
\includegraphics[scale=0.3]{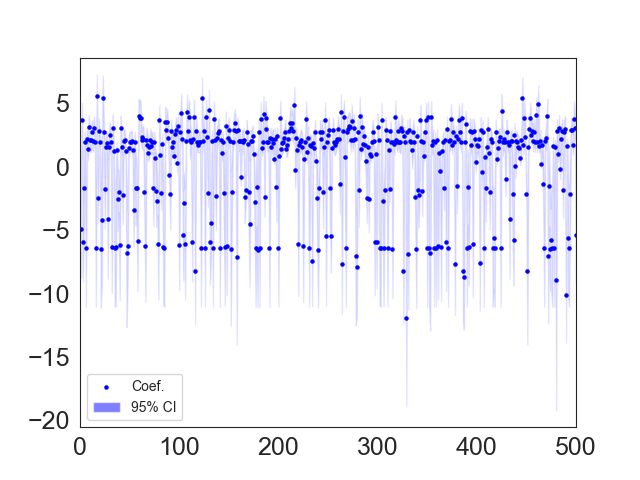}~
\includegraphics[scale=0.3]{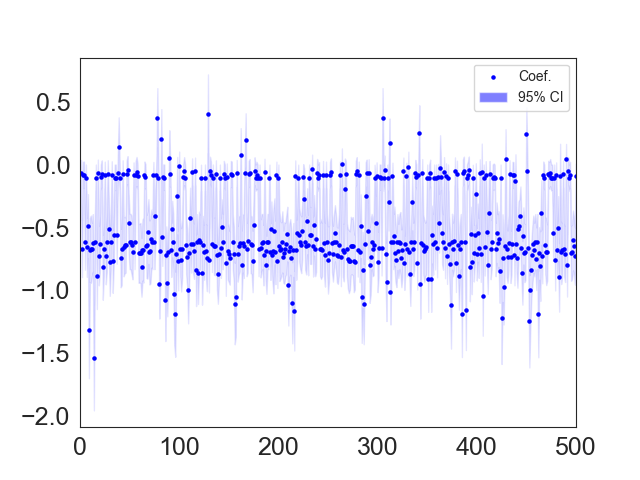}
\end{subfigure}
\caption*{(iii) Mean reversion: Premium (left) and Non-Premium (right)}
\begin{subfigure}[b]{0.5\textwidth}
\includegraphics[scale=0.3]{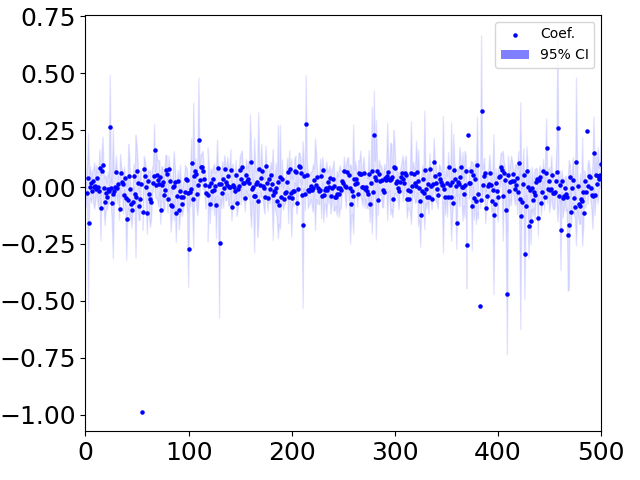}~
\includegraphics[scale=0.3]{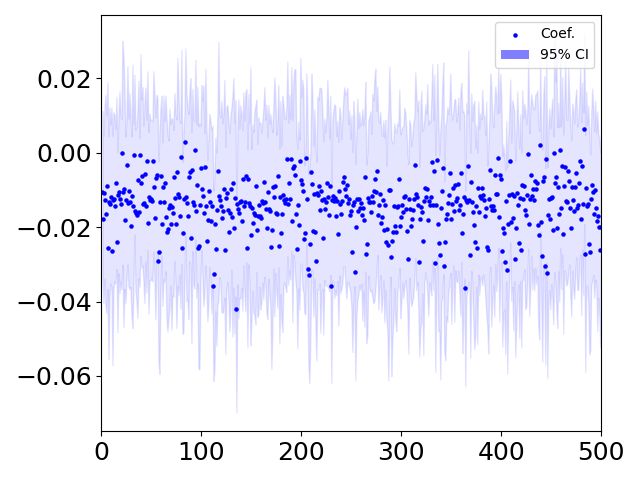}
\end{subfigure}
\caption*{(iv) Prob. of Liquidation: Premium (left) and Non-Premium (right)}
\label{fig:placebo}
\begin{figurenotes}
This figure shows coefficient estimates for premium and non-premium PIPE investors, using randomly generated premium PIPE identities. We randomly pick 34 PIPE investors out of 1,826, label them as premium PIPE investors, and re-estimate our models. We repeat this exercise to have 500 sets of randomly chosen premium PIPE samples. In these five models, the specifications are the same as in Tables \ref{tab:t_redemption_premium} and  \ref{tab:t_r_ann_premium}, column (5), Table \ref{tab:mlot_last} column (3), and Table \ref{tab:relationship_capital_liquidation} columns (6), respectively.
\end{figurenotes}
\end{figure}

We use a placebo exercise to assess whether our choice of premium and non-premium investors is meaningful.  
To this end, we explore the effect of choosing an arbitrary set of PIPE investors and designating them as premium PIPE investors on the estimates. If our results were only noise, then a random group of premium PIPE investors should also give similar estimates. 
To implement this ``placebo'' exercise, we randomly picked 34 out of all 1,826 PIPE investors and labeled them premium investors. Then, we estimate all the models using this new set of premium investors. We repeat this exercise 500 times and present only the coefficients for premium and non-premium investors and the 95\% confidence intervals in Figure \ref{fig:placebo}, in the first and second columns, respectively. The five rows are redemption rate, announcement day return, mean-reversion, liquidation risk, and probability of liquidation.\footnote{For the mean-reversion exercise, we present the first 500 cases with the first-stage F-stat $\geq50$.}

For the redemption rates, comparing the first row in Figure \ref{fig:placebo} with the estimates in Table \ref{tab:t_redemption_premium} column (5), we see that the randomly chosen premium PIPE investors do not affect redemption rates. However, because the non-premium PIPE investors include some (actual) premium PIPE investors, the redemption rate is low but negatively correlated with allocations for non-premium PIPE investors.
Similarly, comparing the second row in Figure \ref{fig:placebo} with the estimates in Table \ref{tab:t_r_ann_premium} column (5), we find no effect of premium PIPE investors on the announcement-day return but a small and positive effect of non-premium PIPE investors on return because the latter set now contains the premium investors, see Table \ref{tab:premium_investor}. These effects are consistent with only the premium investors in Table \ref{tab:premium_investor} producing information.

For the mean-reversion (in the third row), some estimates are negative, consistent with the new premium PIPE investor containing non-premium PIPE investors and vice versa. 
Finally, for the probability of liquidation, comparing rows four in Figure \ref{fig:placebo} with Table \ref{tab:relationship_capital_liquidation} columns (6), we see the estimates for non-premium investors are smaller in magnitude and not statistically significant.

\setcounter{section}{0}
\setcounter{equation}{0}
\setcounter{table}{0}
\renewcommand{\thesection}{B}
\renewcommand{\thesubsection}{B.\arabic{subsection}}
\renewcommand{\thetable}{B\arabic{table}}

\section{Variable Definition\label{section:definition}} 
\begin{table}[h]
\vspace*{-1cm}
\centering
\caption{Variable Definition}
\scalebox{0.66}{
\begin{tabular}{p{15em}p{5em}p{38em}}
\toprule
Variable & Unit & Definition\\
\midrule
\\[-1.8ex]
\multicolumn{3}{l}{\textbf{Panel A: SPAC-Level Characteristics}} \\
\multicolumn{2}{l}{\textbf{SPAC Measures}} \\
IPO proceeds & MM USD & Gross proceeds raised in the IPO, including any full or partial exercise of the Greenshoe.\\
No. of Warrant & & Number of warrants included in the unit issued at the SPAC's IPO. \\
No. of Right & & Number of rights included in the unit issued at the SPAC's IPO. \\
Overallotment & \% & Amount of the overallotment option exercised by the underwriter(s) as a fraction of total IPO proceeds.\\
Listing-day return & \% & SPAC IPO investors' first-day return. \\
Days searching & Day & The number of days between the SPAC's IPO date and the target announcement date.\\
Days remaining & Day & The number of days between the SPAC's target announcement date and the liquidation deadline.\\
Announcement-day return & \% & One-day return on the SPAC's shares on the target's announcement date. \\
Redemption & \% & Redeemed SPAC common shares as a percentage of total shares issued at IPO. \\
Days deSPAC & Day & The number of days between the SPAC's target announcement date and the merger closing date. \\
Cold & pps & The difference between the SPAC's offer price and the SPAC's share price on the closing day of the merger. \\
Liquidation & binary & One if the SPAC liquidated during our sample period, zero otherwise. \\
Target sector & & Sector which the SPAC's registration statement indicates it will focus on in its search for a target for a business combination. \\
Merger sector && Sector in which the merger company operates.\\
Jurisdiction && State (if incorporated in the U.S.) or country in which the SPAC is incorporated at the time of the IPO.\\
No. active SPACs & & Number of SPACs actively searching for targets at the time of the focal SPAC's IPO. \\
No. past successful deals && Number of SPACs that completed the merger process before this focal SPAC's IPO, and initiated by the same sponsor. \\
\\[-1.8ex]
\multicolumn{2}{l}{\textbf{PIPE Participation}} \\
$\mathbbm{1}$\{PIPE (pre)\} & Binary& One if PIPE (pre) investors participate in the deal. \\
$\mathbbm{1}$\{PIPE (at)\}&  Binary& One if PIPE (at) investors participate in the deal. \\
$\mathbbm{1}$\{PIPE (post)\}&  Binary& One if PIPE (post) investors participate in the deal. \\
$\mathbbm{1}$\{Premium PIPE (at)\} &Binary& One if premium PIPE (at) investors participate in the deal. \\
$\mathbbm{1}$\{Non-Premium PIPE (at)\}& Binary& One if non-premium PIPE (at) investors participate in the deal. \\
No. of PIPE (at)        & & Number of PIPE (at) investors.    \\
No. Premium PIPE (at)      && Number of premium PIPE (at) investors.   \\
No. of Non-Premium PIPE (at)   && Number of non-premium PIPE (at) investors. \\

\\[-1.8ex]
\multicolumn{2}{l}{\textbf{PIPE Investment/IPO Proceeds}} \\
\%PIPE (pre) & \%  & PIPE (pre) investors' investment as a percentage of total IPO proceeds. \\
\%Premium PIPE (at) & \% &  Premium PIPE (at) investors' investment as a percentage of total IPO proceeds. \\
\%Non-Premium PIPE (at) & \% & Non-premium PIPE (at) investors' investment as a percentage of total IPO proceeds. \\
\%PIPE (post)    & \% & PIPE (post) investors' investment as a percentage of total IPO proceeds. \\

\midrule
\\[-1.8ex]
\multicolumn{3}{l}{\textbf{Panel B: PIPE-Investor-Level Characteristics}} \\
\\[-1.8ex]
Non-Premium & Binary& One if either the PIPE (at) investor's number of participated PIPE deals or the total commitment amount is not in the top 5th percentile. \\
Investment Amount & MM USD & Dollar amount of the PIPE investor's investment in the SPAC deal.\\
MLOT & MM USD & The difference between the value of securities held by the PIPE investor on the first day of listing of the post-merger company and the PIPE investor's initial investment.\\
lag-MLOT & MM USD & The PIPE investor's MLOT from the previous SPAC deal with the same sponsor. \\
lag$^2$-MLOT & MM USD & The PIPE investor's MLOT from the one before the previous SPAC deal with the same sponsor. \\
MLOT(past) & MM USD & Total MLOT a PIPE investor made in past deals with the sponsor.\\
No. of past SPACs & &Number of SPAC deals participated by the PIPE investor before the focal SPAC deal. \\
Past Investment Amount &MM USD &The PIPE investor's total amount of investments in SPAC deals before the focal SPAC deal. \\
No. of SPACs with other sponsors & &Number of other sponsors' SPAC deals participated in by the PIPE investor before the focal SPAC deal. \\
Past MLOT with other sponsors & MM USD & The PIPE investor's total MLOT made in other sponsors' SPACs before the focal SPAC deal. \\
No. of SPACs with other sponsors, same industry & &Number of other sponsors' SPAC deals that found targets in the same industry as the focal SPAC and participated by the PIPE investor before the focal SPAC deal. \\
Past MLOT with other sponsors, same industry & MM USD & The PIPE investor's total MLOT made in other sponsors' SPACs that found targets in the same industry as the focal SPAC. \\
Outside Option & MM USD & Highest MLOT among all premium PIPE investors not included in the current deal.\\
Participation & Binary & One if the sponsor allocates securities to the non-premium PIPE investor, and zero otherwise.\\
Allocation & &Ratio of the non-premium investor's shares to the total shares allocated to the sponsor's relationship investors.\\
\bottomrule
\end{tabular}}%
\label{tab:def}%
\end{table}%

\end{document}